\documentclass[journal]{IEEEtran}
\ifCLASSINFOpdf
\else
\fi
\newcommand{\etal}{{et al. }}

\usepackage{amssymb,mathrsfs,amsmath}
\usepackage{cite}
\usepackage{graphicx}

\usepackage{float}
\usepackage{tabularx}  
\usepackage{threeparttable}  
\usepackage{multirow}
\usepackage[numbered]{bookmark}
\usepackage{hyperref}
    \hypersetup{
        bookmarksnumbered=true,
        bookmarksopen=true,
        bookmarksopenlevel=5,
        colorlinks=true,
        citecolor=black,
        filecolor=black,
        linkcolor=black,
        urlcolor=black,
        pdfstartview=Fit,
        pdfpagemode=UseOutlines
    }

\begin{document}
%
\title{Recent Advances on Non-Line-of-Sight Imaging: Conventional Physical Models, Deep Learning, and New Scenes}
%
%
%

\author{Ruixu~Geng,
        Yang~Hu,
        and~Yan~Chen,~\IEEEmembership{Senior~Member,~IEEE}
\thanks{R. Geng is with the School of Information and Communication Engineering, University of Electronic Science and Technology of China, Chengdu 611731, China (e-mail: gengruixu@std.uestc.edu.cn).}
\thanks{Y. Hu is with the Department of Electronic Engineering and Information Science, University of Science and Technology of China, Hefei 230000, China (e-mail: eeyhu@ustc.edu.cn)}
\thanks{Y. Chen is with the School of Cyberspace Security, University of Science and Technology of China, Hefei 230000, China (e-mail: eecyan@ustc.edu.cn)}
} 

\maketitle
\begin{abstract}
    As an emerging technology that has attracted huge attention, non-line-of-sight (NLOS) imaging can reconstruct hidden objects by analyzing the diffuse reflection on a relay surface, with broad application prospects in the fields of autonomous driving, medical imaging, and defense. Despite the challenges of low signal-to-noise ratio (SNR) and high ill-posedness, NLOS imaging has been developed rapidly in recent years. Most current NLOS imaging technologies use conventional physical models, constructing imaging models through active or passive illumination and using reconstruction algorithms to restore hidden scenes. Moreover, deep learning algorithms for NLOS imaging have also received much attention recently. This paper presents a comprehensive overview of both conventional and deep learning-based NLOS imaging techniques. Besides, we also survey new proposed NLOS scenes, and discuss the challenges and prospects of existing technologies. Such a survey can help readers have an overview of different types of NLOS imaging, thus expediting the development of seeing around corners.
\end{abstract}


%
\IEEEpeerreviewmaketitle

\begin{figure*}[!h]
    \centering
    \includegraphics[width=1\textwidth]{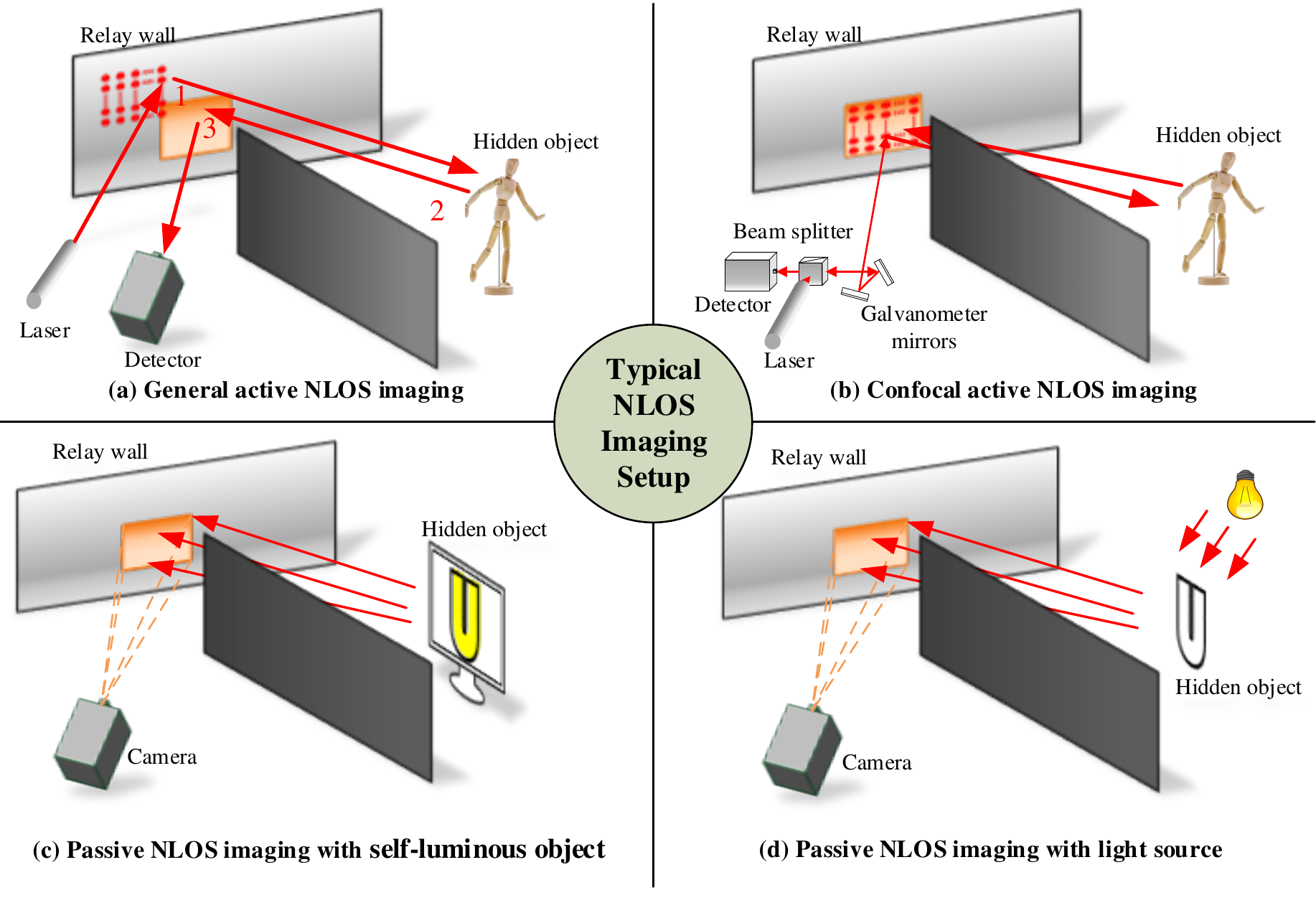}
    \caption{Four typical NLOS imaging setups}
    \label{fig:introFig}
\end{figure*}

\section{Introduction}
\bookmark[dest=\HyperLocalCurrentHref,level=1]{Introduction}
With the rapid development of photon-sensitive sensors and imaging algorithms, optical imaging capabilities have been greatly improved in recent years. Within the line of sight, high-quality imaging can be achieved at a relatively long distance. However, due to the inherent physical constraint of visible light, traditional optical imaging is difficult to see objects outside the line of sight. To break that restriction, non-line-of-sight(NLOS) imaging analyzes the diffuse reflection from a relay wall to image hidden objects, which has broad applications in many fields, such as medical imaging, autonomous driving, and robotic vision~\cite{otooleConfocalNonlineofsightImaging2018,liuVirtualWaveOptics2018}.

According to whether a controllable light source is used, NLOS imaging can be divided into active imaging and passive imaging. Active NLOS imaging often uses expensive external light sources with high temporal resolution(e.g., ultrafast laser) to illuminate the diffuse relay surface. Simultaneously, a sensitive time-resolved detector is used to detect the light reflected on the relay surface, hidden objects, and relay surface in sequence.
The collected effective light is often referred as the three-bounce light since it is reflected by three surfaces (relay surface, hidden objects, and relay surface) in succession. 
Then, the collected three-bounce light is analyzed by different algorithms(e.g., backprojection\cite{veltenRecoveringThreedimensionalShape2012,arellanoFastBackprojectionNonline2017}, inverse methods\cite{heideDiffuseMirrors3D2014} and wave-based methods\cite{liuPhasorFieldDiffraction2020,Lindell:2019:Wave}) to reconstruct the hidden scene.
Since the active methods can collect different kinds of information, including intensity, time, and coherence, the active methods can perform a high-resolution 3D reconstruction. On the other hand, passive methods do not use a controllable external light source but use ambient light or light emitted by hidden objects to complete NLOS imaging. Despite the low cost, passive methods usually can only collect intensity\cite{saundersComputationalPeriscopyOrdinary2019} or limited coherence information\cite{batarsehPassiveSensingCorner2018a,boger-lombardPassiveOpticalTimeofflight2019} and usually complete low-quality 2D reconstruction or localization, while a few recent works can estimate both hidden shape and depth with partial occluder~\cite{saundersMultiDepthComputationalPeriscopy2020,seidelTwoDimensionalNonLineofSightScene2020}.

\begin{figure*}[!h]
    \centering
    \includegraphics[width=1\textwidth]{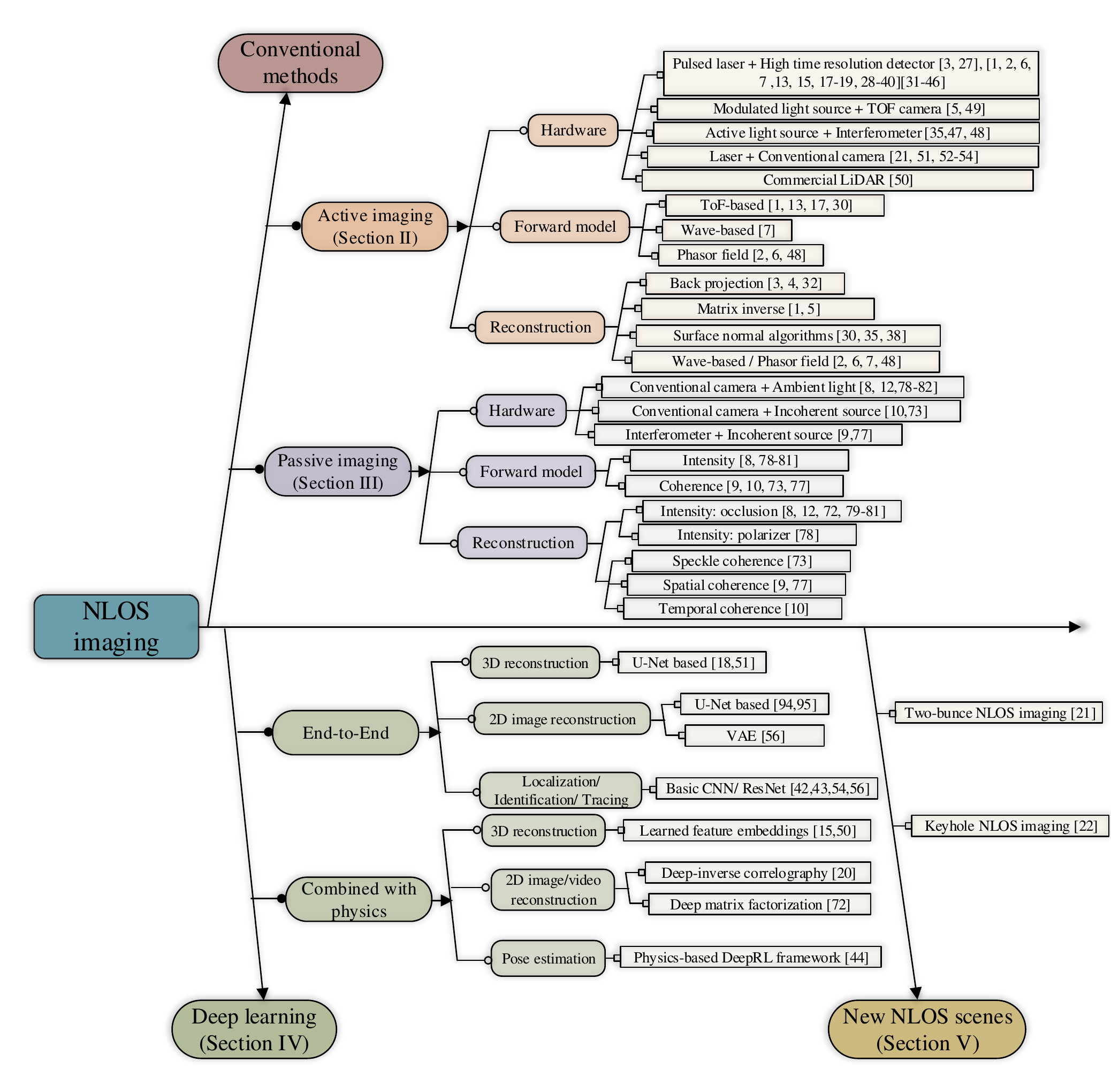}
    \caption{An overview of NLOS imaging.}
    \label{fig:introTree}
\end{figure*}

Four typical NLOS imaging setups are shown in Figure~\ref{fig:introFig}, and there are many challenges when realizing the NLOS imaging. First, due to the high-order loss with distance and environmental noise during the light transport process, NLOS imaging is an ill-posed problem with low SNR, making high-quality reconstruction extremely difficult~\cite{thrampoulidisExploitingOcclusionNonLineofSight2018}. Different hidden scenes may produce the same measurement, which deepens the ill-posedness of the problem~\cite{liuAnalysisFeatureVisibility2019}. Second, the spatial resolution of the reconstruction result is limited by the size of the scanning area (aperture size) and the system's temporal resolution. When the size of the hidden scene is large or complex, the spatial resolution will be limited by the computational complexity~\cite{otooleConfocalNonlineofsightImaging2018}. Third, the data collection time is too long~\cite{yeCompressedSensingActive2021}. Typical time-resolved NLOS imaging needs a scanning process to obtain data, making it difficult to reconstruct in real-time with high quality. Although array detectors (e.g., SPAD array) are promising to eliminate the scanning process~\cite{jinScannerlessNonlineofsightThree2020}, they have not yet been fully explored.

Despite the difficulties, many emerging innovative methods have achieved NLOS imaging under certain scenarios in recent years.
Figure \ref{fig:introTree} summarizes recent NLOS imaging research.
It can be seen that the conventional methods based on physical imaging models, are still the mainstream of research in recent years.
Conventional methods rely on imaging setup and illumination (e.g., active or passive), which are committed to developing three factors in NLOS imaging: advanced hardware systems, accurate forward imaging models and effective reconstruction algorithms. 
For example, \cite{wuNonLineofsightImaging2021} deployed confocal settings and dual-telescope to complete an amazing 1.43km NLOS imaging. Liu \etal exploited the phasor field to convert NLOS imaging model to LOS imaging model\cite{liuVirtualWaveOptics2018,liuPhasorFieldDiffraction2020}, thus achieving high-quality reconstruction of complex scenes. \cite{otooleConfocalNonlineofsightImaging2018} subtly converted NLOS reconstruction into a three-dimensional deconvolution problem. All these works have greatly promoted the development of NLOS imaging.

Besides conventional methods, the application of deep learning in NLOS imaging has also been rapidly developed. According to network design principles, the deep learning methods used in NLOS imaging are divided into end-to-end networks\cite{chopiteDeepNonLineofSightReconstruction2020} and physics-based networks\cite{chen_learned_2020,metzlerDeepinverseCorrelographyRealtime2020}. Compared with conventional algorithms, deep learning algorithms can thoroughly learn the scene prior, automatically extract features, and complete the reconstruction of hidden objects. Although NLOS imaging based on deep learning is still at the early stage, it has broad research prospects for practical applications.
Moreover, some new types of NLOS scenes, such as “imaging behind occluders” \cite{vedaldi_imaging_2020} that used two reflections behind obstacles and keyhole imaging\cite{metzler_keyhole_2021}, have also been proposed, enriching the applications of NLOS imaging. This article aims to make a detailed review of the various types of NLOS imaging research mentioned above, including conventional active and passive methods, deep learning-based NLOS imaging, and new NLOS scenarios. The corresponding challenges and opportunities will also be discussed. We hope this article can help readers have a systematic understanding of existing NLOS imaging research.

Compared to the existing surveys\cite{maedaRecentAdvancesImaging2019,faccioNonlineofsightImaging2020}, this article is much more comprehensive with the coverage of different NLOS imaging scenes, deep learning algorithms, and new NLOS imaging scenarios. As an emerging technology that has developed rapidly in recent years, the algorithms and applicable scenarios of NLOS imaging are constantly growing. Survey~\cite{maedaRecentAdvancesImaging2019} summarized the research of NLOS imaging in detail by classifying existing methods based on ToF information, coherent information, and intensity information. However, it completes the review from the perspective of the information used, and has less introduction to imaging models, reconstruction algorithms, and recent deep learning methods. Besides, in \cite{maedaRecentAdvancesImaging2019}, the passive imaging problem was only a subset of different exploited information and thus not described in detail. Another review~\cite{faccioNonlineofsightImaging2020} provided and summarized the key technologies of laser-based active NLOS imaging, which however lacked the description of passive NLOS imaging, deep learning methods and new NLOS scenes. On the contrary, this article summarizes the latest active and passive NLOS imaging research based on physical methods and provides a detailed summary and analysis of the latest deep learning algorithms, including the advantages, types, challenges, and prospects. Besides, this article also summarizes several new types of NLOS imaging scenarios. Notice that such summarization can help readers have an overall understanding of NLOS imaging, which however cannot be found in the existing surveys~\cite{maedaRecentAdvancesImaging2019, faccioNonlineofsightImaging2020}.

The remainder of this article is organized as follows. In Section \ref{sec2}, we first introduce the existing work of active NLOS imaging from three aspects: the hardware, forward propagation model, and reconstruction algorithms. We also discuss the challenges and prospects of active NLOS imaging. Then, in Section~\ref{sec3}, the related work, including the hardware, forward model, and reconstruction algorithms, as well as the challenges and prospects of passive NLOS methods, are summarized respectively. In Section ~\ref{sec4}, we review the deep learning algorithms that have emerged in recent years from their motivations, network structures, loss functions, corresponding challenges, and prospects. The new NLOS imaging scenarios are discussed in Section~\ref{sec5} and conclusions are drawn finally in Section~\ref{sec6}. 
Figure~\ref{fig:introTree} clearly illustrates the relationship between the structure of this article and the current development of NLOS imaging technologies.
It should be noted that imaging through a scattering medium\cite{lindell_three-dimensional_2020,bertolotti_non-invasive_2012} does not belong to the NLOS imaging scope in this article.

%
%
%
%
\begin{table*}[!t]
    \caption{Active NLOS imaging \label{tab:active}}
    {\begin{tabular}{>{\raggedright}m{3.3cm}m{3.7cm}m{3cm}m{2cm}m{3cm}}
    \hline
    Reference & Illumination & Sensor & Information & Task\\ 
    \hline
    \cite{guptaReconstructionHidden3D2012,veltenRecoveringThreedimensionalShape2012} & Pulsed laser & Streak camera & Time of fight &  3D reconstruction\\
    \cite{buttafavaNonlineofsightImagingUsing2015,laurenzisMultiplereturnSinglephotonCounting2015,heideNonlineofsightImagingPartial2017,jinReconstructionMultipleNonlineofsight2018,mannaErrorBackprojectionAlgorithms2018,otooleConfocalNonlineofsightImaging2018,thrampoulidisExploitingOcclusionNonLineofSight2018,Lindell:2019:Wave,tsaiVolumetricAlbedoSurface2019,pediredlaSNLOSNonlineofsightScanning2019,xinTheoryFermatPaths2019,musarraNonlineofsight3DImaging2019,liuVirtualWaveOptics2018,liuPhasorFieldDiffraction2020,Ahn_2019_ICCV,chopiteDeepNonLineofSightReconstruction2020,Young:2020:dlct,iseringhausen:2018,mannaNonlineofsightimagingUsingDynamic2020,chen_learned_2020,wuNonLineofsightImaging2021,yeCompressedSensingActive2021} & Pulsed laser & SPAD & Time of fight &  3D reconstruction\\
    \cite{chanNonlineofsightTrackingPeople2017a}(long range),\cite{caramazzaNeuralNetworkIdentification2017,musarraDetectionIdentificationTracking2019}(single point) & Pulsed laser & SPAD & Time of fight &  Detection/ Tracking/ Identification\\
    \cite{isogawaOpticalNonLineofSightPhysicsBased2020} & Pulsed laser & SPAD & Time of fight &  Pose estimation\\
    \cite{nam_real-time_2020,jinScannerlessNonlineofsightThree2020} & Pulsed laser & SPAD array & Time of fight &  3D reconstruction\\
    \cite{gariepyDetectionTrackingMoving2016} & Pulsed laser & SPAD array & Time of fight &  Detection/ Tracking/ Identification\\
    \cite{Willomitzer:18,xinTheoryFermatPaths2019,rezaPhasorFieldWaves2019} & Pulsed laser & Interferometer & Coherence &  3D reconstruction\\
    \cite{heideDiffuseMirrors3D2014,kadambiOccludedImagingTimeofflight2016} & Modulated laser & ToF camera & Time of fight &  3D reconstruction\\
    \cite{zhuFastNonlineofsightImaging2021a} & Integrated & LiDAR & Depth and intensity &  3D reconstruction\\
    \cite{chenSteadystateNonLineofSightImaging2019,vedaldi_imaging_2020} & Continuous laser & Conventional camera & Intensity &  3D reconstruction\\
    \cite{kleinTrackingObjectsOutside2016,smithTrackingMultipleObjects2018a,leiDirectObjectRecognition2019} & Continuous laser & Conventional camera & Intensity &  Detection/ Tracking/ Identification\\
    \cite{chandranAdaptiveLightingDataDriven2019} & Incoherent light source (imaging side) & Conventional camera & Intensity &  Detection/ Tracking/ Identification\\
    \cite{DBLP:journals/corr/abs-1810-11710} & Incoherent light source (imaging side) & Conventional camera & Intensity &  2D reconstruction\\
    \cite{Lindell:2019:Acoustic} & Speaker & Microphone & Acoustic "ToF" &  3D reconstruction\\
    \hline
    \end{tabular}}{}
    \end{table*}

\bookmark[dest=\HyperLocalCurrentHref,level=1]{Active Methods}
\section{Active Methods} \label{sec2}
Active NLOS imaging employs a controllable light source (usually a narrow-band laser) and a detector to obtain reflected information, which is then used to reconstruct hidden scenes. This section introduces recent advances of active NLOS imaging from three parts: hardware devices, physical light transport models, and reconstruction algorithms. Besides, the challenges and prospects of active NLOS imaging are analyzed at the end of the section. Table~\ref{tab:active} summarizes active imaging systems, where each row of the table lists the light source, sensor, information, and task for different NLOS imaging techniques.

\bookmark[dest=\HyperLocalCurrentHref,level=2]{Hardware Devices in Active Methods}
\subsection{Hardware Devices in Active Methods}
A variety of detectors, from professional interferometers~\cite{xinTheoryFermatPaths2019} and single-photon counters~\cite{buttafavaNonlineofsightImagingUsing2015,wang2021non} to ordinary cameras and even cell-phone cameras, have been used in NLOS imaging. Different detectors need to be combined with corresponding illumination sources to complete specific NLOS tasks. Here, we introduce such combinations to describe what hardware is used in active NLOS imaging.

\bookmark[dest=\HyperLocalCurrentHref,level=3]{Pulsed laser and high temporal resolution detector}
\subsubsection{Pulsed laser and high temporal resolution detector}

NLOS imaging was first proposed in a time-resolved imaging work by Raskar and  Davis~\cite{raskar5dTimelightTransport2008}, first theoretically evolved by Kirmani \etal~\cite{kirmaniLookingCornerUsing2011} and experimentally demonstrated by Velten \etal~\cite{veltenRecoveringThreedimensionalShape2012}. All of these works~\cite{raskar5dTimelightTransport2008,kirmaniLookingCornerUsing2011,veltenRecoveringThreedimensionalShape2012} were in the context of active time-resolved transient imaging, also called ``light-in-flight imaging'' or ``freezing light in motion''~\cite{faccioNonlineofsightImaging2020}.

Transient imaging uses an ultrafast pulsed laser as the light source and a high time-resolved detector as the camera to measure the time of the photon arrival event in each pulse period, and then obtain the time distribution of photon events through the accumulation of multiple pulse periods, which then is converted to transient images.
Considering that transient imaging is based on pulsed lasers and high-time-resolved detectors, the combination of pulsed laser and high-time-resolved detector is one kind of hardware used in active NLOS imaging.  

\vspace{0.8mm}
\noindent \textbf{Streak cameras}
When NLOS imaging was first experimentally achieved by Velten \etal~\cite{veltenRecoveringThreedimensionalShape2012}, the time-resolved detector was a streak camera with extremely high temporal resolution. As an important optical time characteristic measurement detector, streak camera has been widely used in the experimental research of ultrafast physical processes such as laser fusion and high energy density physics~\cite{shiraga_ultrafast_1997}. 
When the ultrafast light signal passes through the slit of the streak camera, the photoelectric conversion is completed. These photoelectrons are accelerated and focused under the high-voltage electric field's action and enter the scanning system. Within the linear range of the voltage, the vertical distance between the photoelectron's position and the original movement direction is proportional to the time of which the photoelectron enters the scanning plate, hence completing the conversion of time information to space information. Finally, using the function between the optical signal's spatial information and the scanning speed of the streak camera, the temporal information of the optical signal is obtained, which can be reconstructed to get transient images.

The temporal resolution of the streak camera is remarkably high. In~\cite{veltenRecoveringThreedimensionalShape2012}, the theoretical resolution could reach 2ps (limited by a finite temporal-point spread function of the camera, the effective temporal resolution was 15ps), and the most advanced streak camera at present can reach a temporal resolution of $200\sim 300$ fs. However, streak cameras are too expensive (typically more than 70,000\pounds), limiting the practical application of NLOS imaging. Therefore, later works attempt to use an inexpensive time-resolved detector to complete active NLOS imaging.

\vspace{0.8mm}
\noindent \textbf{SPAD (Single Photon Avalanche Diode)}
In 2015, Buttafava \etal~\cite{buttafavaNonlineofsightImagingUsing2015} demonstrated that SPAD (Single Photon Avalanche Diode) is feasible for NLOS imaging.
In Geiger mode, a single photon may trigger an avalanche of about $10^8$ carriers, which can be used as a single photon counter and get pretty accurate photon timing.
The current commercial SPAD covers the wavelength range from $300 nm$ to $1700 nm$. Among them, Si SPAD corresponds to $300\sim 1100 nm$, Ge SPAD corresponds to $800\sim 1600 nm$, and InGaAs SPAD corresponds to $900\sim 1700 nm$\cite{zhou_research_2010}. Therefore, most existing methods with visible light (e.g., $675nm$ in \cite{otooleConfocalNonlineofsightImaging2018}) used Si SPAD (e.g., PDM series from Micro Photon Devices), while the recent $1.43 km$ NLOS imaging \cite{wuNonLineofsightImaging2021} with infrared spectrum ($\sim 1550nm$) used InGaAs/InP negative-feedback SPAD.

Combined with time-correlated single-photon counting technology, SPAD can be used to obtain a histogram of photon arrival time of different points, i.e., transient images. In addition to the relatively lower cost (about 10000\pounds), compared to streak cameras, SPAD has the following advantages. First, as a subset of APDs (avalanche photodiode), although the temporal resolution($\sim 20ps$) is lower than streak cameras, the SPAD has a higher quantum efficiency($\sim 70\%$) which means that it is more suitable for NLOS imaging scenes with weak effective signals. The work in \cite{wuNonLineofsightImaging2021} which used SPAD to complete an amazing $1.43 km$ NLOS imaging is a typical example. 
Moreover, SPAD has been widely used in commercial LiDAR systems, and the SPAD array, which can avoid the mechanical raster scan process, has the potential to save scanning time and realize real-time data collection for active NLOS imaging.

\vspace{0.8mm}
\noindent \textbf{Pulsed laser}
In transient imaging, the laser plays the role of illumination, triggering, and synchronization. Compared with time-resolved detectors, lasers have relatively low requirements and are more affordable than detectors. Nevertheless, it should still meet the following requirements. First, the pulse period of the laser should be adjustable to ensure that it can meet the requirements of different scene sizes without causing excessive noise. Second, in order to achieve accurate synchronization, the pulse width of the laser should be narrow with low jitter. Besides, its wavelength should be consistent with the response frequency of the detector.

Although the combination of pulsed lasers and time-resolved detectors suffer from long scanning time when collecting data, it is still the most popular active NLOS imaging camera method. It can obtain accurate time information and the potential applications of scanning-free technologies, such as SPAD array\cite{jinScannerlessNonlineofsightThree2020,nam_real-time_2020,pei_dynamic_2021}.

\bookmark[dest=\HyperLocalCurrentHref,level=3]{Modulated light source and ToF camera}
\subsubsection{Modulated light source and ToF camera}
In addition to the pulse-based ToF measurement, by encoding the ToF into phase measurement, a ToF camera combined with a modulated light source has also been proposed to obtain the light travel time to complete NLOS imaging\cite{heideDiffuseMirrors3D2014,kadambiOccludedImagingTimeofflight2016}. Compared with pulse-based photon-level detectors, ToF cameras have some obvious advantages. First, it can complete data collection without scanning, thereby reducing data collection time. Second, its price is much lower (about 1000\pounds) than streak camera and SPAD. However, due to the influence of frequency aliasing, the measurement distance of the ToF camera is limited (no more than 10m). Besides, due to the longer exposure time, the ambient noise of the ToF camera is usually greater, limiting the temporal resolution to $ns$ level with $cm$ imaging resolution. 
However, in practical active NLOS imaging scenarios, the modulated light source and ToF camera are on the same side, which would increase the direct bounce signal from the wall and decrease the hidden signal (the third-bounce light), resulting in poor SNR. Therefore, in general, the ToF camera is more suitable for NLOS scenes that require real-time performance without high imaging quality.

\bookmark[dest=\HyperLocalCurrentHref,level=3]{Active light source and interferometry}
\subsubsection{Active light source and interferometry}
Pulse-based detectors and coherent-based ToF cameras are the two main active NLOS imaging cameras. However, their resolution is strongly restricted, difficult to break through the $\sim mm$ level. The resolution of the pulse-based detector is limited by the pulse width of the laser ($\sim ps$), while the coherent ToF camera is limited by the modulation frequency ($\sim ns$). To achieve higher resolution ($\sim \mu m$), interferometers have also been applied to NLOS imaging. Unlike those ToF-based cameras that can only be used for active imaging as mentioned above, the interferometer can not only be combined with narrowband LEDs or coherent light for active imaging, but also be used with ambient light for passive imaging. For active methods, Xin \etal~used the imaging device in~\cite{gkioulekas_micron-scale_2015} to complete the NLOS imaging of coin-sized scenes and achieved femtosecond scale resolution~\cite{xinTheoryFermatPaths2019}. Willomiitzer \etal~utilized lasers with two wavelengths to complete high-resolution NLOS imaging based on superheterodyne interferometry (SHI) with a resolution of about $50 \mu m$~\cite{Willomitzer:18}. Some studies performed passive NLOS imaging by applying narrowband or ambient illumination to hidden objects, which will be introduced in Section ~\ref{sec3}. Compared with streak cameras, SPAD, and ToF cameras, the interferometers can achieve higher resolution but with the disadvantages of higher hardware complexity and cumbersome calibration.

\bookmark[dest=\HyperLocalCurrentHref,level=3]{Laser and conventional camera}
\subsubsection{Laser and conventional camera}
In this paper, a conventional camera refers to a camera that uses conventional intensity sensors, such as CCD or CMOS array, which cannot record spatial/temporal coherent information or ToF information. Since transient images cannot be measured, imaging using traditional cameras is often called steady-state imaging. Because conventional cameras can only record intensity information, they are mainly used in passive NLOS imaging. However, because the conventional camera has the advantages of not requiring scanning and low cost, Chen \etal~completed RGB active NLOS imaging using lasers with different wavelengths for illumination and exploiting conventional cameras to collect intensity information\cite{chenSteadystateNonLineofSightImaging2019}. Due to the lack of distance information, it is not easy to complete high-precision three-dimensional reconstruction only using conventional cameras.

\bookmark[dest=\HyperLocalCurrentHref,level=3]{LiDAR}
\subsubsection{LiDAR}
Although most existing works have used separate light sources and detectors, recent work\cite{zhuFastNonlineofsightImaging2021a} exploited a commercial LiDAR, which integrated light source and detector, to complete NLOS imaging. For the imaging area on the relay surface that is close to LiDAR, most of the collected information is the direct reflection from the relay surface. However, for the area on the relay surface that is closer to the hidden object, the collected information mainly encodes the shape of the hidden object, which can be utilized to restore 3D shapes. 
The existing commercial LiDAR can only provide point cloud output, instead of transient images. Therefore, the classic active NLOS reconstruction methods, such as FBP\cite{veltenRecoveringThreedimensionalShape2012}, LCT\cite{otooleConfocalNonlineofsightImaging2018} and phasor field\cite{liuVirtualWaveOptics2018}, cannot currently be used in LiDAR-based systems. \cite{zhuFastNonlineofsightImaging2021a} used deep learning to fuse point cloud information and reflection intensity information to complete the reconstruction.
Compared with separate laser and time-resolved detectors, integrated LiDAR have lower cost and faster imaging speed, with the cost of poor imaging detail.
To replace the separate laser and cameras in NLOS systems, existing commercial LiDAR needs to solve the problem of crosstalk between multi-channel lasers under NLOS conditions and provide original temporal signals rather than point cloud after processing.

\bookmark[dest=\HyperLocalCurrentHref,level=2]{Forward propagation model}
\subsection{Forward propagation model}
The forward propagation model of active NLOS imaging aims to establish the imaging model of measurement. The current imaging models are mainly divided into ToF-based imaging models and wave-based imaging models. The ToF-based imaging model~\cite{otooleConfocalNonlineofsightImaging2018,veltenRecoveringThreedimensionalShape2012} uses geometric optics to establish the model with the flight distance and surface albedo or normal direction as constraints. The wave-based imaging model~\cite{Lindell:2019:Wave,liuPhasorFieldDiffraction2020} mainly uses wave optics to construct the propagation and reflection of waves with boundary conditions.

\bookmark[dest=\HyperLocalCurrentHref,level=3]{ToF-based model}
\subsubsection{ToF-based model}
The transient-based forward model established using time of flight as a constraint is a classic active NLOS imaging model. This type of model is based on point-by-point scanning, leading to a long scanning time. Therefore, there are many methods proposed to improve the scanning mechanism, represented by the confocal setting~\cite{otooleConfocalNonlineofsightImaging2018}. Besides, 3D reconstruction is an ill-posed problem, and using partial occlusion to add additional constraints is also a representative improvement~\cite{thrampoulidisExploitingOcclusionNonLineofSight2018}. Therefore, we review the general ToF-based NLOS image formation model, followed by the confocal imaging and occlusion-based models.

\vspace{0.8mm}
\noindent \textbf{General NLOS imaging model.}
The imaging model fits the optical transport process by expressing the measurement data $\tau$ as a function of the hidden object space $\Omega \in \mathbb{R}^3$. As shown in Fig.~\ref{fig:introFig}-(a), the emitted laser first irradiates at the illumination point $\mu$ on the wall, causing the first bounce. Then, the second bounce occurs on the surface of the hidden object. Finally, the detection point $\mu'$ on the wall is collected after the third bounce. For a given illumination point $\mu$, the collected  signal $\tau$ by the detection point $\mu'$ at time $t$ can be expressed as~\cite{otooleConfocalNonlineofsightImaging2018}

\begin{equation}
    \begin{split}
    \tau(\mu,\mu',t) =  &\int_{\Omega} \rho(s) F(\mu \rightarrow s \rightarrow \mu')  R(\mu\rightarrow s \rightarrow \mu') \\
    & B(\mu\rightarrow s \rightarrow \mu') ds
    \end{split}
    \label{eq1}
\end{equation}
where $s$ represents a point on the hidden object, and $\rho(s)$ represents the albedo of point $s$. $F$ refers to the optical transfer process from the illumination point $\mu$ to the detection point $\mu'$, reflected at the surface of hidden object $s$. $R$ means the amplitude attenuation. $B$ represents the bidirectional reflectance distribution function (BRDF).

According to geometric optics, with the speed of light $c$, the optical transport process $F$ should satisfy $\|\mu -s\|_{2} + \|\mu' -s\|_{2} = ct$, which can be represented by delta function $\delta()$. 
Please note that this $\delta()$ function is established based on two assumptions: (1) Only three-bounce light is considered; (2) There is no interreflection in the hidden scene.
The attenuation is inversely proportional to the square of the distance. Therefore, assuming that light scatters isotropically, Eq.~(\ref{eq1}) can be re-written as
\begin{equation}
    \begin{split}
    \tau(\mu,\mu',t) =  &\frac{1}{\|\mu -s\|_{2}^{2} \|\mu' -s\|_{2}^{2}}\int_{\Omega} \rho(s) \delta(\|\mu -s\|_{2} \\
    & + \|\mu' -s\|_{2} -ct) ds
    \label{eq1_2}
    \end{split}
\end{equation}
Equation~(\ref{eq1_2}) can be discretized into a linear transform of equations
\begin{equation}
    \tau = \mathrm{A} \rho
    \label{eq2}
\end{equation}
where $\mathrm{A}$ is the optical transport matrix, which maps the hidden scene $\rho$ to measurement $\tau$. In the general NLOS setup, $\rho$ has three spatial dimensions of $x$, $y$, and $z$, and $\tau$ has 5 dimensions (4 spatial dimensions and 1 time dimension). Therefore, the calculation complexity of matrix $\mathrm{A}$ is extremely high. Besides, it is an under-constrained problem inherently. In addition, too many scanning points lead to a long scanning time. In order to solve these problems, many new methods have been proposed (see Sec.~\ref{sec:longDataAcqui} for details). The confocal setup\cite{otooleConfocalNonlineofsightImaging2018} and adding additional obstacles\cite{thrampoulidisExploitingOcclusionNonLineofSight2018,heideNonlineofsightImagingPartial2017} are frequently used methods in recent studies, as introduced in the following.

\vspace{0.8mm}
\noindent \textbf{Confocal NLOS imaging model}
Through a beam splitter, the laser and the detector can be placed coaxially, which means “confocal”, as shown in Fig.~\ref{fig:introFig}-(b). In the confocal NLOS imaging model, the illumination and detection points are same during each scan, i.e., $\mu = \mu'$, which can be substituted into Eq.~(\ref{eq1_2})
\begin{equation}
    \tau(\mu,t) =  \frac{1}{\|\mu -s\|_{2}^{4}} \int_{\Omega} \rho(s) \delta(2\|\mu -s\|_{2} -ct) ds
    \label{eq:confocal}
\end{equation}
O'Toole \etal~proved that under confocal conditions, the imaging model could be further converted to the convolution of the hidden object albedo $\rho$ after resampling with the system response function, which can be solved quickly by Fourier transform. However, under the confocal setting, the SPAD detector receives direct reflections, which may aggravate the pile-up effect~\cite{arlt_study_2013}. In order to alleviate the influence of direct reflections, \cite{wuNonLineofsightImaging2021}~and~\cite{otooleConfocalNonlineofsightImaging2018}~illuminate and image two slightly (to avoid affecting the confocal imaging model) different points on the relay wall. 

\vspace{0.8mm}
\noindent \textbf{Occlusion-based active NLOS imaging model}
For the occlusion-based imaging model, all that needs to be done is to add the visible item $V(\mu,\mu',s)$ to the imaging model, which means the occlusion relationship between $(\mu,s)$ and $(\mu',s)$. Generally, $V(\mu,\mu',s)$ is a boolean variable, i.e., $V(\mu,\mu',s)=0$ if and only if there is no occlusion, otherwise $V(\mu,\mu',s)=1$~\cite{thrampoulidisExploitingOcclusionNonLineofSight2018}. In the study with partial occlusion, the value of $V(\mu,\mu',s)$ is allowed to be continuous~\cite{heideNonlineofsightImagingPartial2017}. An interesting but predictable fact in NLOS imaging is that the occlusion term $V(\mu,\mu',s)$ can be unified into the optical transport matrix $\mathbf{A}$ and reduce the condition number, which is helpful for reconstruction.

It should be noted that Eq.~(\ref{eq1_2}) only considered 3-bounce diffuse reflections. Since the number of photons decreases exponentially with the order of diffuse reflections, 3-bounce reflections is sufficient for current NLOS imaging. The limitations of 3-order reflections are discussed in detail in \cite{liuAnalysisFeatureVisibility2019}.

\bookmark[dest=\HyperLocalCurrentHref,level=3]{Wave-based model}
\subsubsection{Wave-based model} \label{activeAlg}
The wave-based NLOS imaging model regards the forward imaging model as the propagation of waves from hidden objects to detectors in 3D space. Lindell \etal~recorded the light field in space as $\Psi(x,y,z,t)$, then the forward model can be converted to wave propagation from $\Psi(x,y,z,t= 0)$ to $\Psi(x,y,z=0,t)$. 
Specifically, The time-dependent field $\Psi$ can be written as a superposition of plane waves~\cite{Lindell:2019:Wave}
\begin{equation}
    \begin{split}
    \Psi(x, y, z, t)=  &\iiint \Phi\left(k_{x}, k_{y}, k_{z}\right) e^{2 \pi i\left(k_{x} x+k_{y} y+k_{z} z-f t\right)} \\
    & \mathrm{d} k_{x} \mathrm{~d} k_{y} \mathrm{~d} k_{z}
    \label{eqwave_3}
    \end{split}
\end{equation}
where the wave vector $\mathbf{k}=2 \pi \cdot\left(k_{x}, k_{y}, k_{z}\right)$ indicates the direction of propagation of the independent plane wave~\cite{Lindell:2019:Wave}. $f=c \sqrt{k_{x}^{2}+k_{y}^{2}+k_{z}^{2}}$ means the relationship between the wave vector $\mathbf{k}$ and frequency $f$ when the speed of light is $c$. The function $\Phi$ represents the amplitude and phase of each plane wave at $t = 0$. 
After that, f-k migration is used to solve the problem in the frequency domain through fast Fourier transform (FFT) and Stolt interpolation (see Sec.~II.D.\ref{sec:waveBased} for details).

The phasor field methods\cite{rezaPhasorFieldWaves2019,liuPhasorFieldDiffraction2020,liuVirtualWaveOptics2018}, which have attracted widespread attention recently, regard the NLOS imaging as a diffraction-based LOS (line-of-sight) optical imaging problem. The projector function and diffraction function are determined by selecting a suitable LOS template, thereby directly reconstructing the hidden scene. Although based on wave propagation, these methods are all suitable for ToF measurement, making it easy to collect data and apply the model to public NLOS imaging datasets.

\bookmark[dest=\HyperLocalCurrentHref,level=2]{Reconstruction topology}
\subsection{Reconstruction topology}
The reconstruction algorithms of active NLOS imaging have been developed rapidly in recent years\cite{xinTheoryFermatPaths2019,otooleConfocalNonlineofsightImaging2018,liuPhasorFieldDiffraction2020}. Here, we review these reconstruction methods from two aspects: topology and methodology. Reconstruction topology is classified according to different reconstruction targets (Volumetric, Surface and High-level representation), and reconstruction methodology is based on different reconstruction methods (Inverse method and Wave-based methods).

\bookmark[dest=\HyperLocalCurrentHref,level=3]{Volumetric}
\subsubsection{Volumetric}
The volumetric NLOS methods reconstruct the hidden scene by estimating the albedo values of the hidden scenes, which is the most naive idea in NLOS reconstruction. In 2012, Velten \etal~used filter back projection (FBP) to estimate the albedo of the hidden object surface based on the measurement data and completed the NLOS reconstruction for the first time~\cite{veltenRecoveringThreedimensionalShape2012}. Many classic algorithms, such as light cone transform (LCT)~\cite{otooleConfocalNonlineofsightImaging2018}, which exploited confocal settings to convert NLOS reconstruction into a 3D deconvolution problem, and wave-based algorithms~\cite{Lindell:2019:Wave,liuPhasorFieldDiffraction2020}, all aim to restore the albedo of voxels.

The volumetric methods usually have a faster recovery speed (less than 1s for $64\times 64$ spatial resolution~\cite{Young:2020:dlct}). However, because they do not estimate the surface parameters (e.g., normals and BRDFs) of the hidden scene, these methods are not good at recovering details~\cite{tsaiVolumetricAlbedoSurface2019,Young:2020:dlct}.

\bookmark[dest=\HyperLocalCurrentHref,level=3]{Surface}
\subsubsection{Surface}
Unlike the above methods that use volumetric albedo to represent NLOS scenes, a line of work aims to reconstruct the surface of hidden objects.
Such methods usually use special photon measurements (rather than using all photon measurements in volumetric albedo methods~\cite{otooleConfocalNonlineofsightImaging2018,veltenRecoveringThreedimensionalShape2012}) to get detailed geometry.
Tsai \etal~found that the first-returning photons contain the shortest length information to the hidden object, from which the boundary and the surface normal vector of the hidden object can be reconstructed\cite{tsaiGeometryFirstreturningPhotons2017}. Xin \etal~showed that the discontinuities of ToF measurement are produced by special light paths (Fermat Paths), which contain the surface information of the hidden scene\cite{xinTheoryFermatPaths2019}. Since these discontinuities are independent of photon intensity, this approach is robust to different BRDFs.
Besides, by introducing surface normal into the previous transport matrix in Eq.~(\ref{eqK}), the surface normal reconstruction and volumetric albedo reconstruction can be combined to achieve good results~\cite{heideNonlineofsightImagingPartial2017}. 
Similarly, fusing the surface reconstruction into the convolution kernel of the LCT~\cite{otooleConfocalNonlineofsightImaging2018} can also achieve better reconstruction results than using only the volumetric albedo representation~\cite{Young:2020:dlct}.

The above methods restore the surface of the hidden object directly, while other methods adopt the concept of reverse rendering, i.e., to reconstruct the hidden surface by finding the surface parameters (such as BRDF and surface normal) of the hidden object that can fit the measurement data. Considering that accurate reverse rendering methods are time-consuming, a differential renderer can be used to speed up rendering\cite{tsaiVolumetricAlbedoSurface2019}.

\begin{figure*}[!h]
    \centering
    \includegraphics[width=1\textwidth]{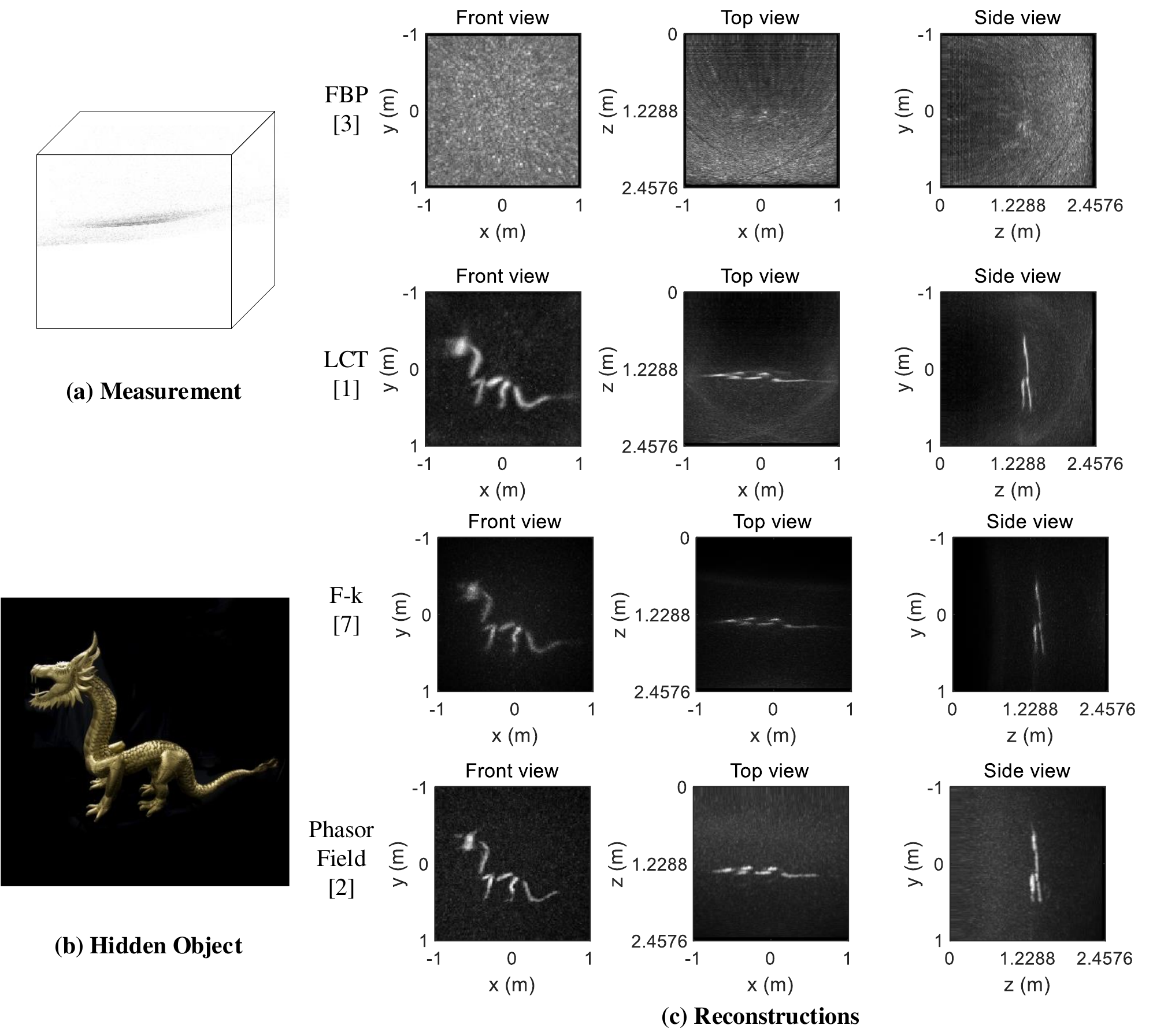}%
    \caption{Reconstructions completed by different methods. (a) Measurement data; (b) Hidden object (dragon); (c) Reconstruction results under several typical methods, including FBP~\cite{veltenRecoveringThreedimensionalShape2012}, LCT~\cite{otooleConfocalNonlineofsightImaging2018}, f-k migration~\cite{Lindell:2019:Wave} and phasor field~\cite{liuVirtualWaveOptics2018}. This figure is finished based on the public code and data of \cite{otooleConfocalNonlineofsightImaging2018,o2017reconstructing,heideNonlineofsightImagingPartial2017,Lindell:2019:Wave}.}
    \label{fig:activeresult}
\end{figure*}

\bookmark[dest=\HyperLocalCurrentHref,level=3]{High-level representation}
\subsubsection{High-level representation}
Besides volumetric and surface-based methods, with the rapid development of deep learning in recent years, some works have exploited high-level representation to complete NLOS reconstruction~\cite{chopiteDeepNonLineofSightReconstruction2020,chen_learned_2020}. Specifically, in a data-driven manner, an encoder is used to extract high-dimensional features of the measurement data, while a decoder is used to map them to the hidden object space to complete the reconstruction. Due to the lack of suitable data sets and the incomplete exploration of the network structure, such methods usually have limited generalization capabilities and reconstruction effects. However, compared to traditional reconstruction methods~\cite{otooleConfocalNonlineofsightImaging2018,liuPhasorFieldDiffraction2020,xinTheoryFermatPaths2019,Young:2020:dlct}, the high-level representation methods based on deep learning have faster inference speed (less than 100ms) and stronger feature extraction capabilities, which is a kind of promising methods.

To emphasize the application prospect of deep learning in NLOS imaging, this paper puts all deep learning methods into Sec.~\ref{sec4}. Therefore, please refer to Sec.~\ref{sec4} for detailed discussion and challenges of the high-level representation methods.

\bookmark[dest=\HyperLocalCurrentHref,level=2]{Reconstruction methodology}
\subsection{Reconstruction methodology}
\label{sec:active_reconstruct}
Active reconstruction algorithms can be divided into two categories according to reconstruction methodology: inverse methods and wave-based (forward) methods.

\bookmark[dest=\HyperLocalCurrentHref,level=3]{Inverse methods}
\subsubsection{Inverse methods}
It can be seen from the forward propagation model (Eq.~\ref{eq1}) that what the active NLOS imaging completed is the inverse process, that is to recover the hidden scene $\rho$ from the measurement $\tau(\mu,\mu',t)$. Under the assumptions of isotropic reflection, no interreflection and only three-bounce light, Eq.~\ref{eq1} degenerates into a linear model, which can be solved by a variety of inverse methods (e.g., back projection~\cite{arellanoFastBackprojectionNonline2017,mannaErrorBackprojectionAlgorithms2018,veltenRecoveringThreedimensionalShape2012} and matrix inverse~\cite{heideDiffuseMirrors3D2014,otooleConfocalNonlineofsightImaging2018}).

\vspace{0.8mm}
\noindent \textbf{Back projection}
For general NLOS imaging scenes, the Dirac function in Eq~(\ref{eq2}) essentially represents an ellipsoidal constraint
\begin{equation}
    r + r_l = c t
\end{equation}

The reconstruction of this model is similar to the ellipsoidal Radon transform used in CT. In confocal NLOS imaging, the ellipsoidal constraint degenerates into a spherical constraint, and the corresponding reconstruction also becomes the spherical Radon transform in CT. Therefore, the back projection algorithm can be directly used to complete the NLOS reconstruction task. In 2012, Velten \etal~used the back projection algorithm to complete the NLOS imaging experiment for the first time\cite{veltenRecoveringThreedimensionalShape2012}. Afterward, researches such as the use of GPU acceleration\cite{arellanoFastBackprojectionNonline2017} and error back projection using iterative algorithms\cite{mannaErrorBackprojectionAlgorithms2018} have also been proposed. Since it has been widely used in CT, the back projection algorithm is simple, easy to understand, and has a low space complexity ($O(N^3)$), which can effectively reconstruct simple hidden scenes. However, the back projection algorithm's time complexity is relatively high ($O(N^5)$), and it is difficult to use other prior information, unable to reconstruct a complex optical transport process.

\vspace{0.8mm}
\noindent \textbf{Matrix Inverse.}
In addition to the back projection algorithm based on the ellipsoidal constraint, NLOS reconstruction can also be completed by directly solving the inverse process of Eq.~(\ref{eq2}), that is
\begin{equation}
    \rho = A^{-1}\{\tau\}
    \label{inverse}
\end{equation}
One straightforward approach is to directly solve the pseudo-inverse of the optical transport matrix $A$, such as using the singular value decomposition. Taking singular value decomposition as an example, since the optical transport matrix $A$ is with the size of $N^3 \times N^3$,  the time and space complexity of the algorithm to find the matrix inversion directly is $O(N^{3\times 3=9})$ and $O(N^6)$ respectively, as discussed in \cite{otooleConfocalNonlineofsightImaging2018}.
Another obvious disadvantage of direct inversion is that it is difficult to use the inherent non-negativity, sparsity, and other priors (see Sec. IV-\ref{sec:PriorsinNLOS} for details) in NLOS imaging, leading to bad results. A common alternative is to solve the following optimization problem\cite{heideDiffuseMirrors3D2014} iteratively
\begin{equation}
\rho_{opt}=\underset{\mathbf{\rho}}{\operatorname{argmin}} \frac{1}{2}\|\mathbf{A} \rho -\tau\|_{2}^{2}+\Gamma(\mathbf{\rho})
\label{eqK}
\end{equation}
Among them, $\Gamma(\mathbf{\rho})$ enables this type of iterative inverse method to utilize multiple constraints. Although iterative methods\cite{guptaReconstructionHidden3D2012,buttafavaNonlineofsightImagingUsing2015} can further reduce errors and achieve good reconstruction results, they require multiple iterations, which is still difficult to apply in real-time. O'Toole \etal~\cite{otooleConfocalNonlineofsightImaging2018}~found that under confocal conditions, the forward imaging process can be expressed in the form of three-dimensional convolution, thus using efficient deconvolution algorithms (such as Wiener filtering) to perform the efficient reconstruction. It reduced the time complexity to $O(N^3log(N))$, which greatly improved the reconstruction speed. 

\vspace{0.8mm}
\noindent \textbf{Inverse Rendering.}
Back projection~\cite{mannaErrorBackprojectionAlgorithms2018,veltenRecoveringThreedimensionalShape2012} and matrix inverse methods~\cite{heideDiffuseMirrors3D2014,otooleConfocalNonlineofsightImaging2018} are both based on voxel representation in reconstruction topology. The voxel-based representation is easy to solve mathematically, but the accuracy is usually low due to ignoring factors such as BRDF and non-Lambert reflections~\cite{tsaiVolumetricAlbedoSurface2019}. On the other hand, by representing the surface of the hidden object, a more accurate forward imaging model (see~\cite{tsaiVolumetricAlbedoSurface2019}) can be established, based on which the optimization methods can be used to solve it to complete the surface reconstruction. Since the imaging models based on surface parameters (such as BRDF and surface normal) rely on rendering, the corresponding reconstruction methods are also referred as inverse rendering, or analysis-by-synthesis. Compared with voxel-based inverse methods, inverse rendering can reconstruct more details, but often requires higher temporal complexity\cite{Young:2020:dlct,heideNonlineofsightImagingPartial2017}.

\bookmark[dest=\HyperLocalCurrentHref,level=3]{Wave-based methods}
\subsubsection{Wave-based methods}
\label{sec:waveBased}
The theoretical basis of the three types of algorithms mentioned above is geometric optics. Besides that, wave-based methods have also achieved rapid development in recent years.
Lindell \etal~described NLOS imaging as a wave propagation problem in three-dimensional space. Inspired by inverse methods used in seismology, it completed NLOS imaging using f-k migration\cite{Lindell:2019:Wave}. In that work, referring $\Psi(x,y,z,t)$ as the field in 3D space as a space-time function, the NLOS reconstruction is converted as 
\begin{equation}
    \Psi(x, y, z=0, t) \quad \Rightarrow \quad \Psi(x, y, z, t=0)
\end{equation}
where $\Psi(x, y, z=0, t)$ and $\Psi(x, y, z, t=0)$ can be regarded as the available measurement data and the hidden target scene, respectively. 
With $t=0$, the functions $\Phi$ and $\Psi$ in Eq.~\ref{eqwave_3} are related by a Fourier transform. By replacing $dk_z$ with $d_f$ and using Stolt interpolation, Eq.~\ref{eqwave_3} can be converted to another representation. In the new representation, when z = 0, functions $\Phi$ and $\Psi$ are again a pair connected by Fourier transform. In this way, taking $\Psi(x, y, z=0, t)$ as input, the hidden scene $\Psi(x, y, z , t=0)$ can be easily reconstructed through three steps: 3D Fourier transform, Stolt interpolation, and inverse 3D Fourier transform. Strictly speaking, f-k migration~\cite{Lindell:2019:Wave} is an inverse method (reconstruction through the reverse process of forward propagation). However, compared to other typical inverse methods such as back projection~\cite{arellanoFastBackprojectionNonline2017,veltenRecoveringThreedimensionalShape2012}, matrix inverse~\cite{heideDiffuseMirrors3D2014} and inverse rendering~\cite{tsaiVolumetricAlbedoSurface2019}, f-k migration does not have an obvious inversion process (e.g., ADMM~\cite{otooleConfocalNonlineofsightImaging2018}), but is more like a forward process completed in the frequency domain.

Besides, Liu~ \etal~\cite{liuPhasorFieldDiffraction2020,liuVirtualWaveOptics2018}~and Reza~\etal~\cite{rezaPhasorFieldWaves2019}~introduced virtual wave phasor to NLOS imaging. Based on the phasor field, NLOS imaging can be transformed into a LOS optical imaging problem and then solved by the existing algorithm in diffraction imaging. 
Different from other methods, phasor field methods~\cite{liuPhasorFieldDiffraction2020,liuVirtualWaveOptics2018} transform the relay wall into a virtual aperture (or lens of any LOS system). The reconstruction is the diffraction integral of the wavefront of the virtual aperture, which is equivalent to the forward propagation process of the measurement data. Therefore, phasor field methods~\cite{liuPhasorFieldDiffraction2020,liuVirtualWaveOptics2018} do not need to reverse the forward process like inverse methods~\cite{otooleConfocalNonlineofsightImaging2018,veltenRecoveringThreedimensionalShape2012} -- they can directly complete the reconstruction through wavefront propagation.

The wave-based approaches have two attractive advantages: (1) It is more robust to the material of the relay surface; (2) It can easily combine NLOS imaging with other related fields, such as LOS imaging and seismic imaging. These advantages, as well as the phase acquisition problem in wave-based approaches, are explained in detail in Faccio \etal's review\cite{faccioNonlineofsightImaging2020}.

It should be noted that not all of the above algorithms are suitable for confocal settings. Specifically, FBP~\cite{arellanoFastBackprojectionNonline2017,veltenRecoveringThreedimensionalShape2012}, phasor field~\cite{liuPhasorFieldDiffraction2020,liuVirtualWaveOptics2018}, and matrix inverse methods~\cite{heideDiffuseMirrors3D2014} are applicable to all NLOS systems. In contrast, LCT~\cite{otooleConfocalNonlineofsightImaging2018} and f-k migration~\cite{Lindell:2019:Wave} are only applicable to confocal NLOS systems. \cite{Lindell:2019:Wave} proposed a conversion method from non-confocal data to confocal data, so that confocal algorithms can also be used to reconstruct non-confocal data. From the perspective of time complexity, for confocal settings, a series of algorithms after LCT~\cite{otooleConfocalNonlineofsightImaging2018} (including f-k migration~\cite{Lindell:2019:Wave}) can reach the time of $O(N^3logN)$ complexity due to the efficient calculation of FFT in the frequency domain, while only the phasor field algorithm~\cite{liuPhasorFieldDiffraction2020} can achieve the same temporal complexity for non-confocal NLOS scenes.
Figure~\ref{fig:activeresult} illustrates the measurement with a dragon as the hidden scene, and the reconstruction results obtained by different reconstruction methods~\cite{otooleConfocalNonlineofsightImaging2018,veltenRecoveringThreedimensionalShape2012,liuVirtualWaveOptics2018,Lindell:2019:Wave}. It can be seen that existing methods can already reconstruct complex hidden scenes. In general, the wave-based methods~\cite{liuVirtualWaveOptics2018,Lindell:2019:Wave} are more suitable for hidden scenes with rich details.

\bookmark[dest=\HyperLocalCurrentHref,level=3]{Detection, location and identification}
\subsubsection{Detection, location and identification}
Under some scenarios, we do not necessarily need to complete the difficult but low-level task of imaging/ reconstruction. On the contrary, some  high-level vision tasks, such as detection, localization, and recognition of hidden objects, are needed. For the detection task, we only need to judge whether there is a peak in the measurement data. 
For localization tasks, a naive but effective algorithm is back projection\cite{chanNonlineofsightTrackingPeople2017a,gariepyDetectionTrackingMoving2016}. Through back projection, the distribution probability of objects in three-dimensional space can be established to determine the object's position. Compared with the reconstruction task, the localization task can greatly reduce the number of scanning points and the resolution of discrete voxels with much less reconstruction time and space complexity. 

For recognition tasks, traditional algorithms are limited by their ability to understand high-level semantics, making it difficult to map from experimental data to label results directly.
Therefore, NLOS recognition is typically  based on data-driven methods (see Sec.~\ref{sec4} for details).

\begin{figure*}[!h]
    \centering
    \includegraphics[width=1\textwidth]{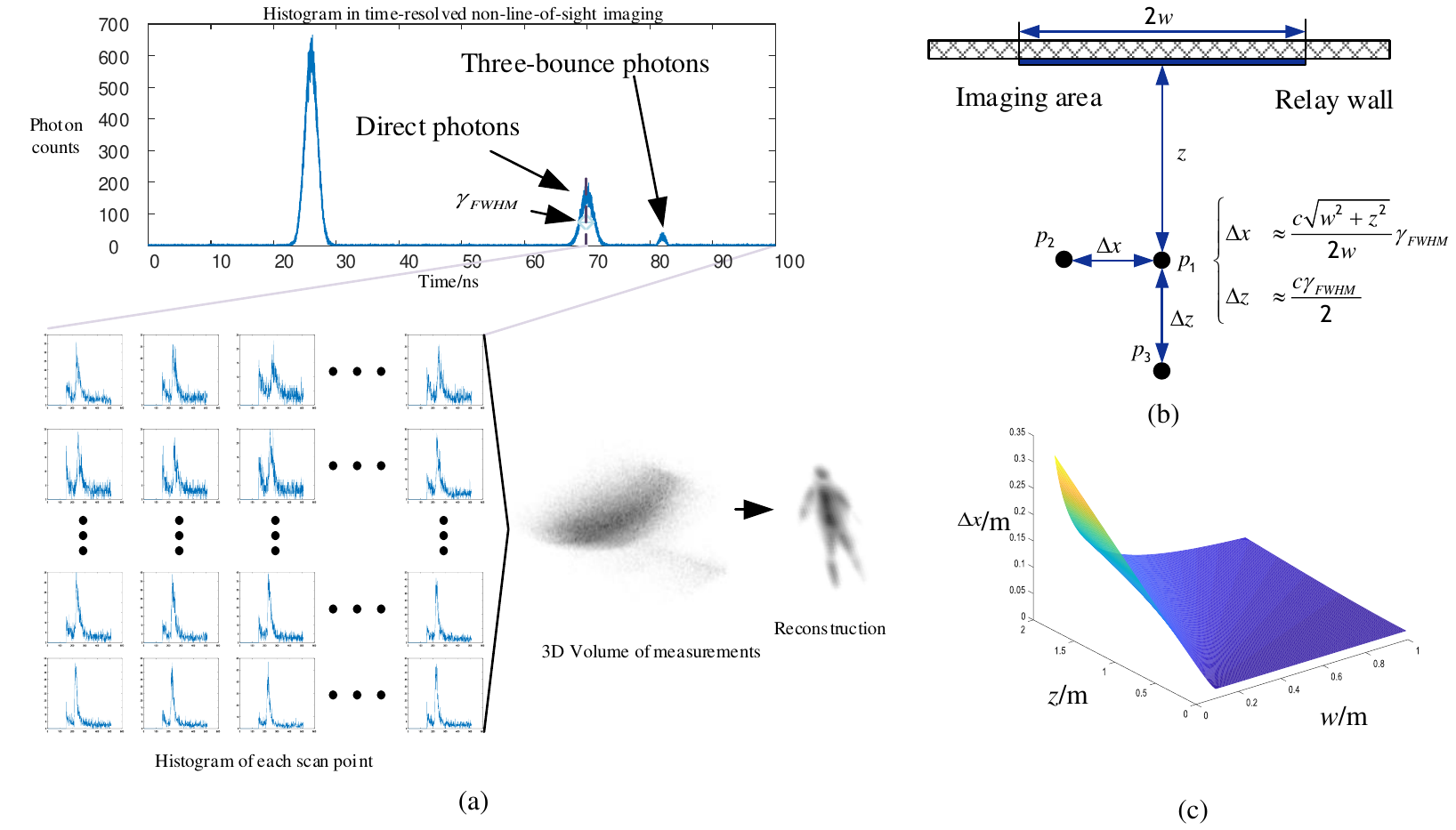}%
    \caption{The principle and spatial resolution of active (confocal) NLOS imaging. (a) The detector receives the photon arrival time histogram at each point, and these histograms form a 3D Volume of measurements $\tau$, which can be used to recover the hidden object (albedo) $\rho$. The measurement data is from \cite{otooleConfocalNonlineofsightImaging2018}. (b) The resolution of the imaging system. (c) The influence of the distance between the object and the wall $z$ and the scan area size $w$ on the horizontal resolution $\Delta x$.}
    \label{fig:active}
\end{figure*}

\bookmark[dest=\HyperLocalCurrentHref,level=2]{Challenges and Prospects}
\subsection{Challenges and Prospects} \label{sec:activeChallenge}
Despite considerable progress in recent years, active NLOS imaging still faces many challenges, including ill-posedness with low SNR, limited resolution, and long data acquisition time. 

\bookmark[dest=\HyperLocalCurrentHref,level=3]{Ill-posedness with low SNR}
\subsubsection{Ill-posedness with low SNR}
Since the collected effective signal is three-bounce of reflected light, and the signal intensity has an attenuation of $r^2\sim r^4$\cite{otooleConfocalNonlineofsightImaging2018} (varies with different reflective materials) with distance $r$, the signal strength is weak. In many NLOS scenes (such as long-distance imaging\cite{wuNonLineofsightImaging2021,liSinglephotonImaging2002021}), the echo can reach the order of a single photon. On the other hand, there are various kinds of noise. All of the dark count and after-pulse of SPAD, the pile-up effect of TCSPC, and the ambient light can cause noise. Therefore, NLOS imaging tasks have a very low SNR, which makes NLOS imaging very challenging. Moreover, hidden objects in different locations may contribute the same measurement value\cite{liuAnalysisFeatureVisibility2019}, which further exacerbates the problem's ill-posedness.

The methods of improving the SNR have been discussed in \cite{liSinglephotonImaging2002021,wuNonLineofsightImaging2021,Li:20}, including improving the receiving device to increase the detection efficiency and thus the signal strength, combining  spatial filtering with multimode fiber, spectral filtering with narrow-band filters and temporal filtering through gate-mode SPAD, and using polarizers to minimize noise. Based on the above methods, an amazing 1.43km NLOS imaging can be achieved. In addition to improving the signal-to-noise ratio from the hardware, it is also important to improve the reconstruction algorithm to alleviate the ill-posedness through various prior constraints, such as the sparsity, non-negativity, surface normal, partial occlusion (as discussed in Sec.~\ref{activeAlg}) and the recently developed data-driven scene prior (see in Sec.~\ref{sec4}).

\bookmark[dest=\HyperLocalCurrentHref,level=3]{Long data acquisition time}
\subsubsection{Long data acquisition time}
\label{sec:longDataAcqui}
Active NLOS imaging requires a full raster scan of a relay wall. This point-by-point scanning mechanism leads to long data acquisition time, which hinders real-time NLOS imaging applications. Current work usually uses a multifunction I/O device (e.g., NI DAG USB-6343) to control the galvanometer mirrors (e.g., Thorlabs GVS012) to scan point by point and completes the event synchronization with the photon counter(e.g., TCSPC, PicoHarp 300). Limited by the scanning speed of the galvanometer and the minimum number of photons at each point, the scanning frequency is only 1Hz to 10Hz (depending on the imaging conditions) in state-of-the-art NLOS systems, such as LCT\cite{otooleConfocalNonlineofsightImaging2018} and f-k migration~\cite{Lindell:2019:Wave}. When the number of scanning points is $64\times 64$, the scanning speed of less than 10Hz is difficult to complete real-time data acquisition.

\begin{figure*}[!h]
    \centering
    \includegraphics[width=1\textwidth]{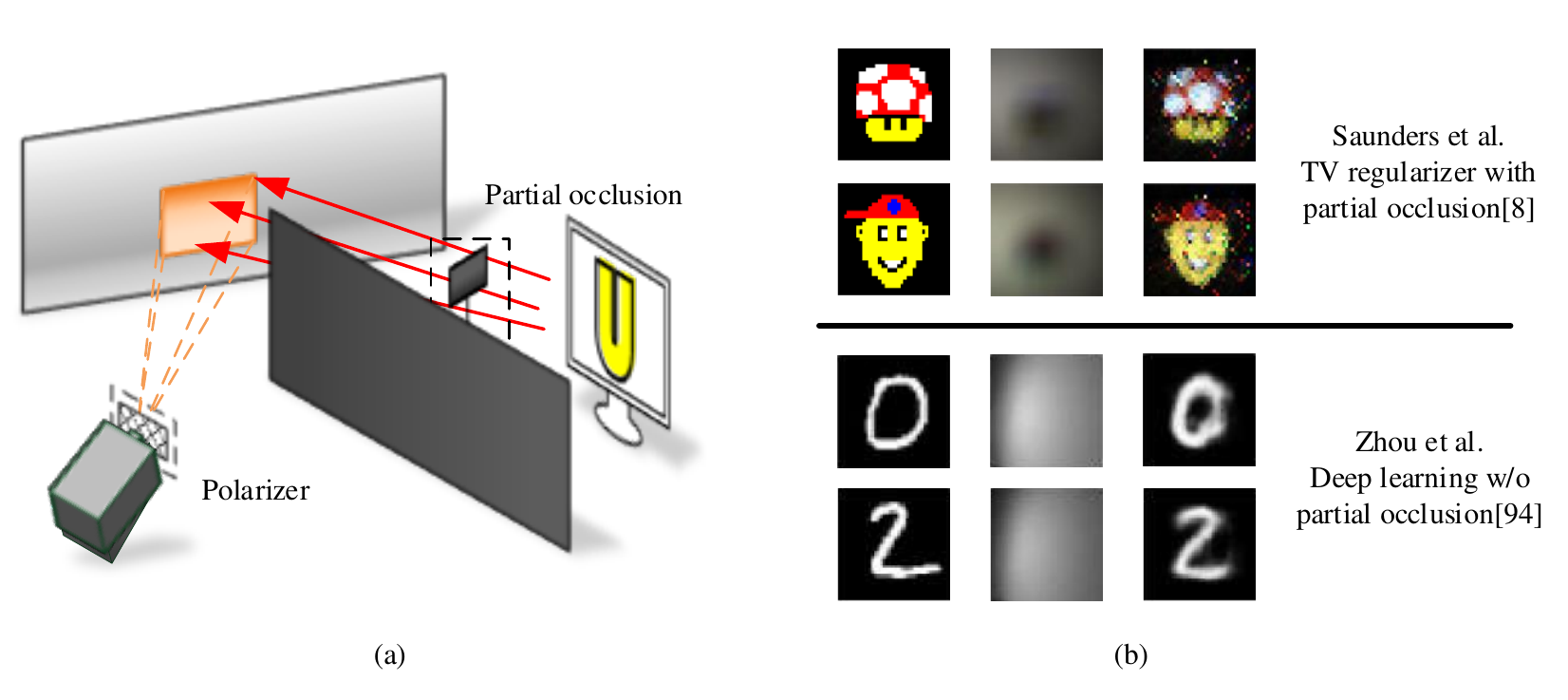}%
    \caption{Passive NLOS imaging: (a) Common constraints in passive NLOS imaging; (b) Passive NLOS imaging results.}
    \label{fig:passive}
 \end{figure*}

\vspace{0.8mm}
\noindent \textbf{Reducing scanning points.}
In theory, each scan point in NLOS imaging contains the whole hidden object's shape information. Therefore, although reducing the scan points would inevitably reduce the imaging quality, it is possible only to scan a few points to complete the imaging task. Liu \etal~studied the impact of randomly removing some test points on the reconstruction results~\cite{liuRoleWignerDistribution2020}. Ye \etal~applied compressed sensing to NLOS imaging, and only $5\times 5$ points can achieve a resolution of $64\times 64$,  thereby greatly reducing the time required for scanning~\cite{yeCompressedSensingActive2021}. Isogawa \etal~further changed the scanning methods, using the circular scanning path, and proposed $C^2NLOS$(Circular and confocal NLOS)~\cite{isogawaEfficientNonLineofSightImaging}. $C^2NLOS$ improved the limit of the scanning speed of the galvanometer through circular scanning and reduced the number of scanning points, thereby reducing the data acquisition time.

\vspace{0.8mm}
\noindent \textbf{SPAD array.}
In addition to reducing the scanning points, the development of SPAD array to achieve scannerless NLOS imaging is a very promising direction. There have been the latest researches that develop algorithms suitable for SPAD array~\cite{liuPhasorFieldDiffraction2020,nam_real-time_2020} or directly use  $32\times 32$ SPAD camera to complete scannerless NLOS imaging~\cite{jinScannerlessNonlineofsightThree2020}.
Multiple lasers simultaneously illuminating would cause crosstalk problems, which is unacceptable. Therefore, the current NLOS imaging systems based on SPAD array all adopt the structure of "one laser, multiple pixels". In essence, this is equivalent to multiple parallel conventional non-confocal NLOS imaging systems. Therefore, any algorithm suitable for non-confocal reconstruction, including but not limited to back projection~\cite{arellanoFastBackprojectionNonline2017}, f-k migration~\cite{Lindell:2019:Wave} and phasor field, can be used in NLOS imaging systems based on SPAD array.~\cite{jinScannerlessNonlineofsightThree2020} verified that the reconstruction quality using commercial SPAD array based on phasor field~\cite{liuPhasorFieldDiffraction2020} is close to the early single-pixel NLOS imaging~\cite{veltenRecoveringThreedimensionalShape2012}.

\vspace{0.8mm}
\noindent \textbf{Steady-state imaging.}
Another possible alternative is to replace the pulsed laser with a steady-state laser~\cite{chenSteadystateNonLineofSightImaging2019} or even an ordinary condensing light source (such as a projector)~\cite{chandranAdaptiveLightingDataDriven2019} and use a conventional camera to collect the imaging area directly. These steady-state methods usually rely on data-driven deep learning methods for reconstruction, which will be discussed in Sec.~\ref{sec4}.

\bookmark[dest=\HyperLocalCurrentHref, level=3]{Limited resolution}
\subsubsection{Limited resolution}
Like LiDAR and Radar, the resolution of NLOS imaging is also divided into axial resolution $\Delta z$ and transverse resolution $\Delta x$. Time jitter $\gamma$, the time domain response of the imaging system greatly influences $\Delta x$, and it also determines $\Delta z$. $\gamma$ mainly depends on the detector's temporal resolution, laser pulse width, laser spatial divergence, and other components. In the recent long-distance NLOS imaging research~\cite{wuNonLineofsightImaging2021}, the time jitter $\gamma$ has been formulated. Another factor of $\Delta x$ is the size of the scanning area (
equivalent to the size of the aperture). Usually, we use $w$ to represent the radius of the scanning area, as shown in Fig.~\ref{fig:active}-(b). For a confocal system, the resolution of the system in two directions $\Delta z$ and $\Delta x$ are~\cite{otooleConfocalNonlineofsightImaging2018}

\begin{equation}
    \begin{cases}
    \Delta x& \approx \frac{c \sqrt{w^{2}+z^{2}}}{2 w} \gamma_{FWHM}\\
    \Delta z& \approx \frac{c \gamma_{FWHM}}{2}
    \end{cases}
    \label{res}
\end{equation}

Among them, $\gamma_{FWHM}$ refers to the full width at half maximum (FWHM) of the time jitter $\gamma$, as illustrated in Fig.~\ref{fig:active}-(a). From Fig.~\ref{fig:active}-(c), it can be seen that reducing the time jitter $\gamma$ and increasing the size of the scanning area $w$ are effective ways to improve the confocal system resolution.
For other active NLOS scenes, \cite{kadambiOccludedImagingTimeofflight2016,liuAnalysisFeatureVisibility2019,buttafavaNonlineofsightImagingUsing2015} discussed the limits of spatial resolution. \cite{liuAnalysisFeatureVisibility2019} also analyzed how the position and the normal direction of the hidden object affect the imaging results from the Fourier domain.

\bookmark[dest=\HyperLocalCurrentHref,level=1]{Passive Methods}
\section{Passive Methods} \label{sec3}
Passive NLOS imaging aims to see hidden scenes without using a controllable external light source, which is a very challenging problem. This section has a similar structure with Sec.~\ref{sec2}. It reviews passive NLOS imaging from three aspects: data acquisition devices, physical models, and reconstruction algorithms. Finally, we discuss the challenges and prospects of passive NLOS imaging.
Table~\ref{tab:passive} summarizes the existing passive NLOS imaging technologies, as well as their lighting conditions, sensors, available information, and target tasks.

\bookmark[dest=\HyperLocalCurrentHref,level=2]{Cameras in passive methods}
\subsection{Cameras in passive methods} \label{sec31}

\bookmark[dest=\HyperLocalCurrentHref,level=3]{Conventional camera}
\subsubsection{Conventional camera}
In the absence of a controllable light source, the light field remains steady. Therefore, the most useful information is intensity, which can be recorded by conventional cameras, i.e.,  charge-coupled device (CCD) and complementary metal-oxide-semiconductor (CMOS). Most passive NLOS imaging research used conventional cameras to collect data. However, their illumination was uncontrollable and different, including ambient light\cite{saundersComputationalPeriscopyOrdinary2019,aittalaComputationalMirrorsBlind2019} and active incoherent light\cite{katz2014non}\footnote{Although active illumination is used, it is still considered as passive NLOS imaging because the light source is on the side of the hidden object and is uncontrollable.}.
When using ambient light, the hidden scene produces shadows (or speckles) on the wall. The only information available at this time is the intensity information of the shadows. However, if a narrowband spatially-incoherent source is used for illumination, the obtained speckles can encode the hidden scene's information, which can be reconstructed by the optical memory effect\cite{fengCorrelationsFluctuationsCoherent1988,freundMemoryEffectsPropagation1988,osnabruggeGeneralizedOpticalMemory2017}. 

Conventional cameras are the most inexpensive but have at least two limitations among all the cameras discussed in this article. First, the shadow is very sensitive to ambient light intensity, i.e., SNR decreases as the ambient light intensity increases. Second, for broadband illumination, the optical memory effect is no longer valid, due to which the problem is extremely ill-posed and usually requires additional constraints and priors.

\bookmark[dest=\HyperLocalCurrentHref,level=3]{Interferometer}
\subsubsection{Interferometer}
Considering the limitations of intensity information, some studies employ interferometry to obtain phase information and use coherence to solve it, as illustrated in Tab.~\ref{tab:passive}. 
Batarseh~\etal~used the Dual-Phase Sagnac Interferometer (DuPSaI) to measure the spatial coherence of the light field~\cite{batarsehPassiveSensingCorner2018a}, thereby completing the detection and positioning of the hidden broadband ambient light source. Beckus~\etal~combined spatial coherence and intensity information to complete multi-modal passive NLOS imaging~\cite{beckusMultiModalNonLineofSightPassive2019}. In addition to spatial coherence, temporal coherence can also be used to obtain depth information. Boger-Lombard~\etal~utilized a diffuser and an ordinary camera to provide passive ToF information through the time coherence of the measurement~\cite{boger-lombardPassiveOpticalTimeofflight2019}. 
The interferometer enables depth information measurement in passive NLOS scenes and can be fused with intensity information to improve imaging quality. However, it requires a precise and complex calibration process, and the depth information it contains is very limited, which can only be used for discrete scenes.

\bookmark[dest=\HyperLocalCurrentHref,level=2]{Physical model}
\subsection{Forward imaging model}
Considering the classic intensity-based passive NLOS imaging scene, as shown in Fig.~\ref{fig:introFig}-(c), the goal is to reconstruct the hidden scene based on the speckle area's intensity information $d$ on the wall. Assuming that each point on the hidden scene is an independent point light source, the measured intensity information is

\begin{equation}
   I_{\tau}(d)=\int\int_{s\in S} A(s,d)I_{\rho}(s)ds
\end{equation}
Here $ I_ {\tau}$ is the observed projection image of resolution $p\times q$, $I_{y}(\tau)$ is the light intensity on the projection area $d$. $ I_ {\rho}$ is the hidden target image of resolution $m \times n$ displayed on the screen, and $I_{\rho}(s)$ is the intensity of the pixel point $s$ of $I_x$. Besides, $A(s,d)$ is the optical transport from the point light source $s$ to area $d$ on the relay wall, and $S$ denotes all pixels on the whole screen. The model can be discretized as
\begin{equation}
   \mathbf{I_{\tau}} = \mathbf{A}\mathbf{I_{\rho}} + \mathbf{N}
   \label{eq2_1}
\end{equation}
where $\mathbf{N}$ is an additional background term.

It can be seen that Eq.~(\ref{eq2}) and Eq.~(\ref{eq2_1}) look very similar. Because passive NLOS imaging usually only collects intensity information, the condition number of $\mathbf{A}$ in Eq.~(\ref{eq2_1}) is larger, meaning more ill-posed. 
In addition to the above-mentioned intensity-based forward model, imaging models based on other principles and settings (including the Fresnel model\cite{beckusMultiModalNonLineofSightPassive2019}, partial occlusion\cite{saundersComputationalPeriscopyOrdinary2019,seidelTwoDimensionalNonLineofSightScene2020}, spatial coherence\cite{batarsehPassiveSensingCorner2018a,beckusMultiModalNonLineofSightPassive2019}, and polarizers\cite{tanakaPolarizedNonLineofSightImaging2020}) can all be discretized to inverse optimization problems similar with Eq.~(\ref{eq2_1}). 
Passive methods based on speckle coherence have a special imaging model and reconstruction process, which will be explained in the following (see Eq.~(\ref{eqK_2})).

\begin{table*}[!t]
   \caption{Passive NLOS imaging \label{tab:passive}}
   {\begin{tabular}{m{2.5cm}m{3.5cm}m{3cm}m{4cm}m{3cm}}
   \hline
   Ref & Illumination & Sensor & Information and constraints & Task\\ 
   \hline

   \cite{boger-lombardPassiveOpticalTimeofflight2019} & Incoherent light source (object side) & Conventional camera & Temporal coherence &  Detection/ Tracking/ Identification\\
   \cite{katz2014non} & Incoherent light source (object side) & Conventional camera & Speckle coherence &  2D reconstruction\\
   \cite{batarsehPassiveSensingCorner2018a} & Incoherent light source (object side) & Interferometer & Spatial coherence &  Detection/ Tracking/ Identification\\
   \cite{beckusMultiModalNonLineofSightPassive2019} & Ambient light & Interferometer + Conventional camera & Spatial coherence + Intensity&  2D reconstruction\\
   \cite{yedidiaUsingUnknownOccluders2019,torralbaAccidentalPinholePinspeck2012,aittalaComputationalMirrorsBlind2019,DBLP:conf/iccp/SeidelMMSFYG19,saundersComputationalPeriscopyOrdinary2019,seidelTwoDimensionalNonLineofSightScene2020} & Ambient light & Conventional camera & Intensity with partial occluder &  2D reconstruction\\
   \cite{tanakaPolarizedNonLineofSightImaging2020} & Ambient light & Conventional camera & Intensity with polarizer &  2D reconstruction\\
   \cite{boumanTurningCornersCameras2017} & Ambient light & Conventional camera & Intensity of moving object &  2D reconstruction\\
   \cite{boumanTurningCornersCameras2017} & Ambient light & Conventional camera & Intensity &  Detection/ Tracking/ Identification\\
   \cite{8747343} & Infrared radiation & Infrared camera & Infrared intensity &  Detection/ Tracking/ Identification\\

   \hline
   \end{tabular}}{}
   \end{table*}

\bookmark[dest=\HyperLocalCurrentHref,level=2]{Reconstruction with constraints}
\subsection{Reconstruction with constraints} \label{memoryEffect1}
Passive NLOS imaging is extremely ill-posed, and it is difficult to complete high-quality reconstruction only with conventional cameras. Existing researches reduce the condition number of the transport matrix $\mathbf{A}$ by adding additional constraints, thereby improving the imaging quality, as shown in Tab.~\ref{tab:passive}. This section will discuss reconstruction algorithms under three common constraints, including partial occluder\cite{saundersComputationalPeriscopyOrdinary2019,yedidiaUsingUnknownOccluders2019}, coherence\cite{beckusMultiModalNonLineofSightPassive2019} and polarizer\cite{tanakaPolarizedNonLineofSightImaging2020}.
Data-driven scene priors are another common constraint, which will be explained in Sec.~\ref{sec4}.

\bookmark[dest=\HyperLocalCurrentHref,level=3]{Partial occluder}
\subsubsection{Partial occluder}
In the most primitive cameras, large-area obstructions can be used to form pinholes to tighten the beam and complete imaging\cite{newhallHistoryPhotography1982}. In other early computational imaging applications, such as stereo vision\cite{andersonRolePartialOcclusion1994}, light field recovery\cite{baradadInferringLightFields2018a,veeraraghavanDappledPhotographyMask2007} and image synthesis\cite{10.1145/1073204.1073257}, partial occlusion also played an important role. 
In the field of passive NLOS imaging, many state-of-the-art works also exploit partial occlusion. Among them, some tasks need to know or estimate the prior knowledge of occlusion (such as position and shape) before they can reconstruct hidden scenes using the prior as constraints\cite{saundersComputationalPeriscopyOrdinary2019}. Others only utilized partial occlusion to improve the conditioning of the problem without knowing its specific information\cite{torralbaAccidentalPinholePinspeck2012,yedidiaUsingUnknownOccluders2019,aittalaComputationalMirrorsBlind2019,seidelTwoDimensionalNonLineofSightScene2020,boumanTurningCornersCameras2017}, as illustrated in Fig~\ref{fig:passive}-(a).

Passive NLOS imaging modelled by Eq.~(\ref{eq2_1}) can be solved as an optimal problem
\begin{equation}
\mathbf{I_{\rho,opt}}=\underset{\mathbf{I_{\rho}}}{\operatorname{argmin}} \frac{1}{2}\|\mathbf{A} \mathbf{I_{\rho}} -\mathbf{I_{\tau}}\|_{2}^{2}+\Gamma(\mathbf{I_{\rho}})
\label{eqK_2}
\end{equation}
where $\Gamma$ represents regularization terms, such as total variation (TV) or other sparsity constraints. When the position of the partial occlusion $p$ is known, the optical transport matrix $\mathbf{A}$ can be estimated, after which the inverse problem can be solved through Eq.~(\ref{eqK_2}). The results of \cite{saundersComputationalPeriscopyOrdinary2019}, a representative work using TV regularization, are shown in Fig.~\ref{fig:passive}-(b).
When the partially occluded position $p$ cannot be obtained, the constraints at different times can also be used to estimate the optical transport matrix $\mathbf{A}$ and the corresponding hidden scene $I_{\rho}$. For example, Aittala~\etal~performed matrix decomposition through an unsupervised learning method to complete passive NLOS imaging\cite{aittalaComputationalMirrorsBlind2019}. This is essentially similar similar to blind deconvolution\cite{chanTotalVariationBlind1998,kundurBlindImageDeconvolution1996,levinUnderstandingEvaluatingBlind2009}, as discussed in \cite{yedidiaUsingUnknownOccluders2019}.

\bookmark[dest=\HyperLocalCurrentHref,level=3]{Polarizer}
\subsubsection{Polarizer}
Although partial occlusion is inevitable in many NLOS scenes, adding occlusion may still change the hidden scene. However, in practical applications, it is not the hidden scene but the imaging system that can be changed. 
A demonstrated alternative method is to add a polarizer in front of the camera, as shown in Fig.~\ref{fig:passive}-(a). In this case, a small angle difference between the light paths leads to a large intensity change, and the optical transport matrix $\mathbf{A}$ becomes\cite{tanakaPolarizedNonLineofSightImaging2020}
$$
A^{\prime}(s, d)=A(s, d) \lambda\left(\boldsymbol{\omega}_{\mathbf{i}}, \boldsymbol{\omega}_{\mathbf{o}}, \mathbf{p}\right)
$$
where $\lambda$ indicates the light leaking/blocking effect caused by the polarizer. $\omega_\mathbf{i}$,$\omega_\mathbf{o}$ and $\mathbf{p}$ indicate viewing vector, incident vector and the polarizer axis, respectively. $\lambda$ decreases the condition number of matrix $\mathbf{A}$, thereby improving the image quality.

\bookmark[dest=\HyperLocalCurrentHref,level=3]{Coherence}
\subsubsection{Coherence} \label{memoryEffect}
Three types of coherence, speckle coherence, spatial coherence and time coherence, have been used in passive non-line-of-sight imaging. Among them, speckle coherence does not require a special interferometer, but needs a narrowband incoherent light source as illumination. 
The speckle coherence model is based on the optical memory effect\cite{freundMemoryEffectsPropagation1988}, which completes NLOS imaging through the scattering medium and corner objects by analyzing the coherence information between hidden object intensity and measured speckle
\begin{equation}
   [\mathbf{I_{\tau}} \star \mathbf{I_{\tau}}](\theta)= [\mathbf{I_{\rho}} \star \mathbf{I_{\rho}}](\theta)
   \label{eq2_2}
\end{equation}
where $\star$ represents the autocorrelation operation. Using the phase recovery algorithm\cite{fienupPhaseRetrievalAlgorithms1982,bertolottiNoninvasiveImagingOpaque2012a}, hidden object's diffraction-limited image $\mathbf{I_{\tau}}$ can be recovered from its autocorrelation\cite{katz2014non}. The field of view of speckle autocorrelation imaging depends on the range of memory effect, which is generally very small. This is the bottleneck that limits the speckle correlation based on the memory effect.

Reconstruction algorithms based on spatial and temporal coherence are no longer subject to the viewing angle limitation caused by the optical memory effect\cite{freundMemoryEffectsPropagation1988}, but require a very sensitive and costly interferometer. Since the interference effect is sensitive to the light source's number and size, spatial coherence and temporal coherence can usually only be used to complete the positioning, detection of discrete points, and very simple shape restoration. The combination of coherence and intensity information can effectively overcome this limitation. For example, multi-modal passive NLOS imaging that combines intensity information and spatial coherence can achieve promising results\cite{beckusMultiModalNonLineofSightPassive2019}.

\bookmark[dest=\HyperLocalCurrentHref,level=3]{Deep Learning}
\subsubsection{Deep Learning} \label{deepLearning}
Data-driven scene priors are another common constraint, which can be exploited by deep learning methods. Although data-driven passive methods~\cite{DBLP:journals/corr/abs-1810-11710,zhouNonlineofsightImagingPhong2020,yuNonLineofSightImagingDeep2019} also belong to passive NLOS imaging, this article will explain it in Sec.~\ref{sec4} together with active data-driven NLOS methods to highlight the application of deep learning in this field.
Besides, in Sec.~\ref{sec4}, unsupervised passive NLOS imaging algorithms~\cite{aittalaComputationalMirrorsBlind2019} based on physical constraints (matrix factorization) are also introduced.

\bookmark[dest=\HyperLocalCurrentHref,level=2]{Challenges and prospects}
\subsection{Challenges and prospects}
Because there is no controllable light source, passive NLOS imaging can only obtain intensity information and limited coherent information, leading to low reconstruction quality. The two main challenges for passive NLOS imaging are high ill-posedness and limited measurement.

\bookmark[dest=\HyperLocalCurrentHref, level=3]{Ill-posedness}
\subsubsection{Ill-posedness}
Passive NLOS imaging is an extremely ill-posed problem. The reasons can be summarized as follows. (1) The collected data contains little effective information. Since most of the information collected by passive NLOS imaging is intensity information, it can be regarded as a projection of a hidden scene on a two-dimensional plane, without any available temporal information. (2) There is no effective space coding. The problem of passive NLOS imaging is somewhat similar to the recent lensless imaging, but there are no encoders such as moiré fringes\cite{tajima_lensless_2017}, optical phased array\cite{fatemi_88_2017} and known scattering medium\cite{sahoo_single-shot_2017}, which results in the data collected on diffuse reflection surfaces (without valid BRDF encoding) being messy and the degree of ill-posedness being high. (3) The influence of ambient light. When there is ambient light, it is difficult to separate the ambient light from the effective speckle, which also leads to an increase in the ill-posedness\cite{maedaRecentAdvancesImaging2019}.

\vspace{0.8em}
\noindent\textbf{Adding constraints.} 
Adding constraints is an effective way to improve the condition. When the additional constrained parameters are unknown, accurate estimation of constrained parameters (such as the position and shape of partial occlusion) can also improve the imaging effect. Besides partial occlusion~\cite{saundersComputationalPeriscopyOrdinary2019,yedidiaUsingUnknownOccluders2019} and polarizer~\cite{tanakaPolarizedNonLineofSightImaging2020} discussed above, the data-driven methods have a good application prospect, which will be discussed in Sec.~\ref{sec4}.

\begin{table*}[!t]
   \caption{NLOS imaging based on deep learning\label{tab:deeplearning}}
   {\begin{tabular}{>{\raggedright}m{1.5cm}m{2.3cm}m{2cm}m{2.7cm}m{4cm}m{3cm}}
   \hline
   Ref & Network & Input & Output & NLOS Scene& Train set\\ 
   \hline
   \cite{zhouNonlineofsightImagingPhong2020} & End-to-End: U-Net & Scattering image & 2D Image &  Ambient light + conventional camera like \cite{saundersComputationalPeriscopyOrdinary2019} & Experimental and synthetic data\\
   \cite{yuNonLineofSightImagingDeep2019} & End-to-End: a variant of the U-Net & Speckle image & 2D Image &  Incoherent light source + conventional camera like \cite{katz2014non} & Experimental data\\
   \cite{chenSteadystateNonLineofSightImaging2019} & End-to-End: a variant of the U-Net & Scattering image & 3D reconstruction &  CW laser + conventional camera & Synthetic data\\
   \cite{chopiteDeepNonLineofSightReconstruction2020} & End-to-End: a variant of the U-Net & Transient images & 3D reconstruction &  pulsed laser + SPAD like \cite{otooleConfocalNonlineofsightImaging2018} & Synthetic data\\
   \cite{DBLP:journals/corr/abs-1810-11710} & End-to-End: CNN/ VAE & Scattering image & 2D Localization/ identification/ reconstruction &  Incoherent light source + conventional camera like \cite{chandranAdaptiveLightingDataDriven2019} & Synthetic data\\
   \cite{musarraDetectionIdentificationTracking2019,caramazzaNeuralNetworkIdentification2017} & End-to-End: CNN & Temporal histogram & Detection/ Identification/ Tracing &  Pulsed laser + SPAD like \cite{otooleConfocalNonlineofsightImaging2018} & Experimental data\\ 
   \cite{leiDirectObjectRecognition2019} & End-to-End: ResNet & Steady image & Identification &  CW laser + conventional camera like \cite{chenSteadystateNonLineofSightImaging2019} & Synthetic data\\ 
   \cite{aittalaComputationalMirrorsBlind2019} & Deep matrix factorization & Scattering image/ video & 2D image/ video &  Ambient light + conventional camera like \cite{saundersComputationalPeriscopyOrdinary2019} & Experimental data\\
   \cite{metzlerDeepinverseCorrelographyRealtime2020} & Deep-inverse correlography & Speckle image & 2D Image &  CW laser + conventional camera like \cite{chenSteadystateNonLineofSightImaging2019} & Synthetic data\\
   \cite{chen_learned_2020} & Learned feature embeddings & Transient images & 3D reconstruction &  Pulsed laser + SPAD like \cite{otooleConfocalNonlineofsightImaging2018} & Synthetic data\\
   \cite{zhuFastNonlineofsightImaging2021a} & Learned feature embeddings & Depth and intensity images & 3D reconstruction &  Commercial LiDAR & Synthetic data\\
   \cite{isogawaOpticalNonLineofSightPhysicsBased2020} & Physics-based deepRL framework & Transient images & Pose estimation &  Pulsed laser + SPAD like \cite{otooleConfocalNonlineofsightImaging2018} & Synthetic data\\ 

   \hline
   \end{tabular}}{}
   \end{table*} 

\bookmark[dest=\HyperLocalCurrentHref, level=3]{Limited measurement}
\subsubsection{Limited measurement} 
In practical application scenes, adding appropriate constraints between the hidden object and the observation wall may not be allowed, which leads to the limitation of passive NLOS measurement. Therefore, passive NLOS imaging only exploiting intensity information is extremely challenging~\cite{saundersComputationalPeriscopyOrdinary2019}. Existing research can obtain coherence through the interferometer, which contains some depth information but is limited by the shape and size of hidden objects and cannot complete high-quality reconstruction.

\vspace{0.8em}
\noindent\textbf{Multimodal measurements fusion}
Beckus~\etal~combined spatial coherence information and intensity information to complete multimodal passive NLOS imaging, which provides a path to improve the imaging quality\cite{beckusMultiModalNonLineofSightPassive2019}.
The basic principle of passive NLOS imaging is to collect visible light emitted/reflected from hidden objects. However, in addition to visible light, hidden objects also emit/reflect electromagnetic waves on other wavelengths, such as infrared ($\sim 780nm$) and radio signals everywhere in the environment. On the other hand, NLOS imaging based on infrared wave\cite{8747343} and radio\cite{he_non-line--sight_2020} has also been explored in recent years. Therefore, if visible light can be fused with other types of electromagnetic waves to obtain multimodal measurement information, it will hopefully break through the bottleneck of passive NLOS imaging and achieve high-quality reconstruction without occlusion and scene constraint.

\bookmark[dest=\HyperLocalCurrentHref,level=1]{Deep Learning Methods}
\section{Deep Learning Methods} \label{sec4}
Conventional physics-based reconstruction algorithms have been discussed above. In recent years, with the successful application of data-driven algorithms (e.g., deep learning\cite{lecunDeepLearning2015}) in blind deconvolution \cite{kupynDeblurGANBlindMotion2018}, depth estimation\cite{eigen_depth_2014}, and LOS imaging\cite{lindellSinglephoton3DImaging2018,su_deep_2018,marco_deeptof_2017}, recent works have also explored the possibility of deep learning in NLOS imaging. This section first explains the significance of using deep learning algorithms from a priori perspective, then introduces the end-to-end learning algorithms, as well as the hybrid algorithms that combine deep learning with physical models. Finally, we discuss the challenges and prospects of deep learning algorithms. 
Table~\ref{tab:deeplearning} summarizes the existing deep learning-based NLOS imaging technologies and their network, input, output, and training datasets.

\bookmark[dest=\HyperLocalCurrentHref,level=2]{Priors in NLOS imaging}
\subsection{Priors in NLOS imaging}
\label{sec:PriorsinNLOS}
Some priors have been effectively used in NLOS imaging, as reviewed below.

\vspace{0.8em}
\noindent\textbf{Non-negative prior.}
The non-negative prior includes two parts. First, for the optical transport matrix $\mathbf{A}$, by definition, the value of each element is non-negative; Second, for the reconstruction model based on the albedo $\rho$, the albedo of each voxel is also non-negative. The non-negativity of the light transport matrix and albedo $\rho$ together constitute non-negative prior.

\vspace{0.8em}
\noindent\textbf{Sparsity prior.}
Since the measurement data only contains the surface information of the hidden scene, which is sparse in the entire 3D space, the albedo (target scene) $\rho$ has a sparsity prior. A common sparse constraint is to directly use the $L_1$-norm constrain, i.e., $\|\rho\|_{1}$, and \cite{heideDiffuseMirrors3D2014} added different weights to each voxel. Another sparse constraint is a gradient constraint, which limits the gradient in the depth axis is sparse. The last type of sparsity constraint is that there should be only one non-zero value on the z-axis corresponding to each $(x, y)$ coordinate\cite{heideDiffuseMirrors3D2014}. Most NLOS reconstruction algorithms used sparse constraints as the main component of the regularization term $\Gamma(\rho)$\cite{heideNonlineofsightImagingPartial2017,Young:2020:dlct}.

\vspace{0.8em}
\noindent\textbf{Total variational (TV) prior.}
TV represents the sum of discrete gradients, and the TV regularization term is widely used in edge-preserving denoising\cite{rudin_nonlinear_1992,chambolle_introduction_2010} and sparse reconstruction of low-sample data(e.g., compressed sensing)\cite{tang_performance_2009}. For 3D NLOS reconstruction, TV prior represents the sparsity of hidden objects, especially sparsity on the z-axis, which is the gradient constraint discussed above. For two-dimensional NLOS imaging, the TV prior is mainly to restore the image from the inverse problem and maintain the edge information. Due to the effectiveness of the total variation constraint, many NLOS imaging studies have used the TV prior as the regularization term\cite{saundersComputationalPeriscopyOrdinary2019,wuNonLineofsightImaging2021,metzlerDeepinverseCorrelographyRealtime2020,tanakaPolarizedNonLineofSightImaging2020}.

\vspace{0.8em}
\noindent\textbf{Other priors.}
The above three types of priors apply to most NLOS imaging problems. Besides, there are some priors suitable for specific NLOS imaging. Specifically, for ToF-based NLOS imaging, the ellipsoid constraint\cite{otooleConfocalNonlineofsightImaging2018,veltenRecoveringThreedimensionalShape2012}, as well as the surface parameter constraint (e.g., the discontinuity of the ToF measurement contains the information of the surface normal)\cite{xinTheoryFermatPaths2019,tsaiGeometryFirstreturningPhotons2017} are two important priors. For NLOS imaging with partial occluders, the prior knowledge of the occluder's position and shape are also significant for high-quality reconstruction\cite{yedidiaUsingUnknownOccluders2019, thrampoulidisExploitingOcclusionNonLineofSight2018}. Besides, the BRDF prior also plays an important role in improving imaging results\cite{kadambiOccludedImagingTimeofflight2016}.

\bookmark[dest=\HyperLocalCurrentHref,level=3]{Unexploited prior -- scene prior}
\subsubsection{Unexploited prior -- scene prior}
However, there are still some priors that have not been effectively used. Therefore, using this undeveloped prior information is a meaningful way to improve NLOS imaging quality.

Scene priors is an important kind of unexploited priors, which includes both low-level features (e.g., smoothness) and high-level features (e.g., semantic information) of hidden scenes. Scene priors are particularly important for specific reconstruction tasks, such as NLOS imaging in autonomous vehicles or medical imaging.
Compared with other priors, most scene priors are implicit and difficult to extract manually. Therefore, conventional physics-based methods are challenging to formulate and efficiently utilize these scene priors.

The data-driven methods based on deep learning can learn the input data distribution, thereby efficiently completing the mapping from the input space to the output space, which has been utilized in a wide range of applications in recent years. For NLOS imaging, if there is enough reliable data, the data-driven methods can learn the scene prior effectively and improve reconstruction quality. Moreover, although the data-driven methods take a long time in the training phase and require high-performance computing hardware, it can almost achieve real-time processing in the test phase, which helps improve the real-time performance of NLOS imaging.

\begin{figure*}[!h]
    \centering
    \includegraphics[width=1\textwidth]{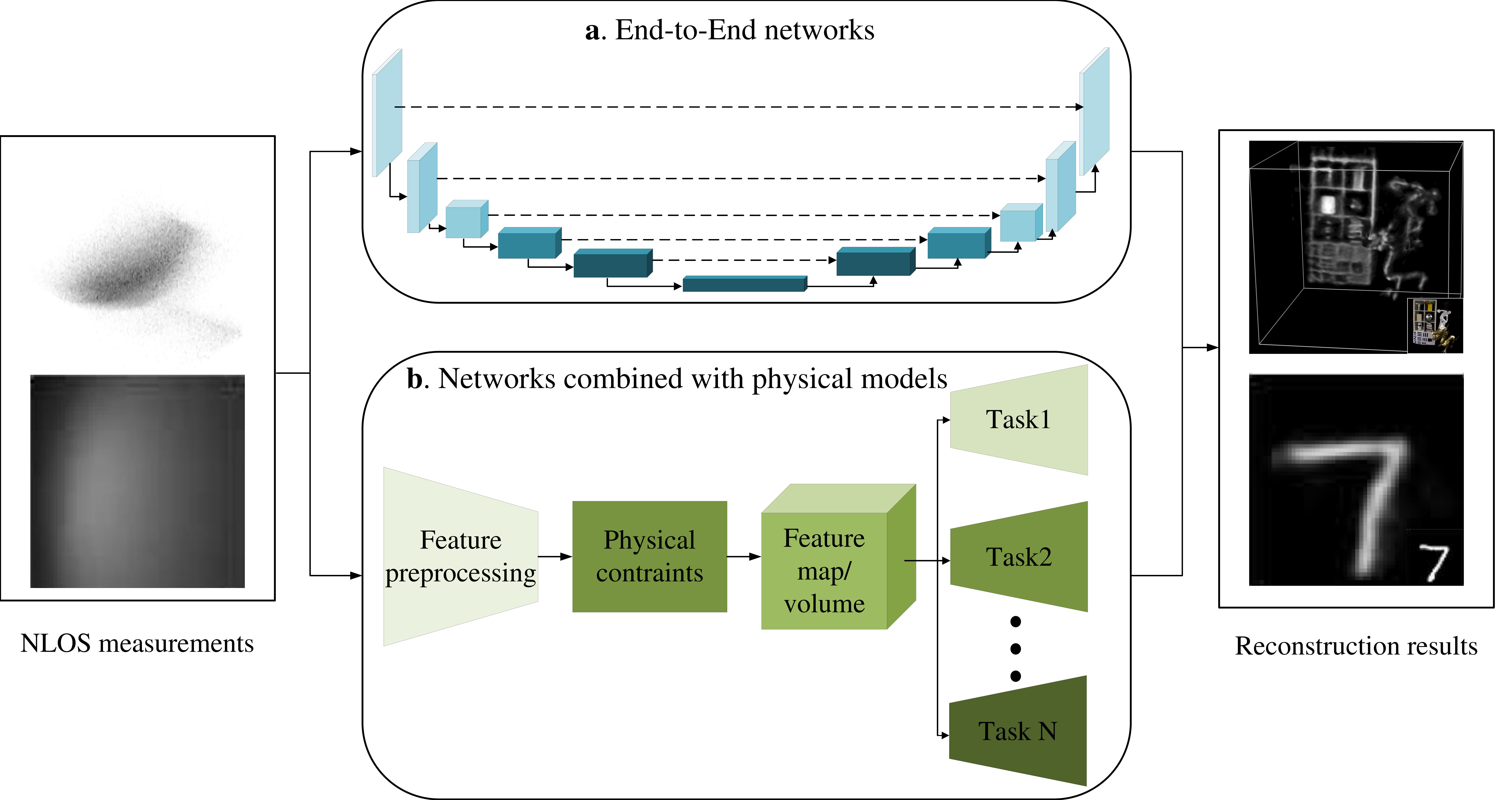}%
    \caption{Two types of models of NLOS imaging based on deep learning: (a) End-to-End Networks. (b) Physics-based Models. The reconstruction results in \cite{chen_learned_2020} and \cite{zhouNonlineofsightImagingPhong2020} are used as a demonstration.}
    \label{fig:deeplearning}
\end{figure*}

Limited by datasets and generalization, data-driven NLOS imaging research is still in the early stage. In recent years, several studies utilized deep learning to propose innovative reconstruction algorithms. The existing works can be roughly divided into two categories. One is to exploit the powerful representation ability of deep learning to build an end-to-end neural network directly (Fig.~\ref{fig:deeplearning}-(a)). The other is to combine the advantages of deep learning with physical models' constraints to improve imaging performance (Fig.~\ref{fig:deeplearning}-(b)).

\bookmark[dest=\HyperLocalCurrentHref,level=2]{End-to-End algorithms}
\subsection{End-to-End algorithms} \label{endToEnd}
NLOS imaging can be expressed as an inverse problem, as shown in Eq.~(\ref{inverse}). Therefore, an end-to-end deep learning model can be built, where the input is the measurement (e.g., $\tau$), and the output is the reconstruction of the hidden scene (e.g., $\hat{\rho}$). Then, a loss function is used to measure the difference between the output $\hat{\rho}$ and ground truth $\rho$, followed by back-propagation used to optimize the network weights, and finally, a good reconstruction result can be obtained. 

Most of the existed NLOS imaging works based on deep learning used the end-to-end network structure, as reviewed in Tab.~\ref{tab:deeplearning}. Chen \etal~used conventional cameras and continuous laser illumination to collect steady-state NLOS imaging data and corresponding hidden scenes~\cite{chenSteadystateNonLineofSightImaging2019}. After that, a U-Net~\cite{ronnebergerUnetConvolutionalNetworks2015a}~based network was trained to complete the mapping from the measurement to the corresponding hidden scenes. In \cite{chenSteadystateNonLineofSightImaging2019}, the loss function was a multi-scale (for different resolutions) $L_2$ loss function.
Similarly, \cite{zhouNonlineofsightImagingPhong2020} and \cite{yuNonLineofSightImagingDeep2019} used the U-Net based network structure, with cross-entropy and ordinary $L_2$ norm as loss function to complete the passive NLOS imaging in Fig.~\ref{fig:introFig}-(c) and (d). They achieved better imaging performance than conventional methods\cite{tanakaPolarizedNonLineofSightImaging2020,saundersComputationalPeriscopyOrdinary2019}, as shown in Fig.~\ref{fig:passive}-(b).
For ToF-based transient NLOS imaging scenes, Chopite \etal~synthesized a large number of training images through rendering and noise models, and modified part of the input and output in U-Net from 2D tensor to 3D tensor using $L_2$ loss\cite{chopiteDeepNonLineofSightReconstruction2020}. Finally, based on the end-to-end deep neural network, \cite{chopiteDeepNonLineofSightReconstruction2020} completed the mapping from transient measurement to depth map.

In addition to reconstruction tasks, recognition is also an important goal in NLOS scenes. Due to the limited capability of conventional algorithms in high-level semantics, NLOS recognition mainly uses data-driven end-to-end algorithms based on deep learning. Lei \etal~completed the recognition of MNIST~\cite{lecunGradientbasedLearningApplied1998a} and human posture under different NLOS settings~\cite{leiDirectObjectRecognition2019}. \cite{caramazzaNeuralNetworkIdentification2017} showed that the deep neural network can classify and recognize hidden scenes with only one SPAD scanning point. In addition, a recent study~\cite{isogawaOpticalNonLineofSightPhysicsBased2020} has combined LSTM~\cite{hochreiterLongShortTermMemory1997} and physical models to complete 3D human pose estimation. Besides, Satat \etal~used deep neural networks to complete object classification through scattering media\cite{satatObjectClassificationScattering2017}.
For passive NLOS scenes, Tancik \etal~utilized CNN for positioning and recognition, achieving a high recognition accuracy\cite{DBLP:journals/corr/abs-1810-11710}. Besides, \cite{DBLP:journals/corr/abs-1810-11710} also explored the use of variational autoencoder (VAE)\cite{kingma_auto-encoding_2014} to complete NLOS reconstruction.

\bookmark[dest=\HyperLocalCurrentHref,level=2]{Network combined with physical models}
\subsection{Network combined with physical models} \label{combined}
The end-to-end method is convenient for training, but it is highly dependent on the dataset. When the dataset scale is large enough, end-to-end learning can learn the mapping well to achieve good results. However, end-to-end learning may have problems such as over-fitting and poor generalization ability when the data size is insufficient. Due to the cumbersome data acquisition process, NLOS imaging lacks real large-scale datasets (most of the existing large-scale datasets are synthetic). Hence, some state-of-the-art studies in recent years proposed a reconstruction algorithm combining deep learning and physical models instead of the end-to-end network only, as shown in Tab.~\ref{tab:deeplearning}.

Metzler \etal~used the optical memory effect (see Sec.~\ref{sec3}-\ref{memoryEffect1}) to complete NLOS imaging~\cite{metzlerDeepinverseCorrelographyRealtime2020}. Instead of using conventional phase recovery algorithms (e.g., hybrid input-output (HIO)~\cite{fienupPhaseRetrievalAlgorithms1982} and alternating minimization (Alt-Min)\cite{netrapalli_phase_2015}), \cite{metzlerDeepinverseCorrelographyRealtime2020}~adopted U-Net to complete phase recovery, with the $L_1$ norm after autocorrelation as loss function.
Recently, Chen \etal~proposed an innovative NLOS imaging network structure rather than using the well-known U-Net like other works~\cite{chen_learned_2020}. It first embedded the synthesized transient images into a feature space (feature embedding), then propagated to the hidden volume. After that, the network is divided into several parts with clear physical meaning, such as visibility network, image rendering, and depth estimation. To some extent, the physical model constrains the deep network to have good generalization ability. Although~\cite{chen_learned_2020} contains multiple components, it can still conveniently run in an end-to-end mode during training.
Another example of exploiting deep learning to promote the development of NLOS imaging is LiDAR-based imaging~\cite{zhuFastNonlineofsightImaging2021a}.
For traditional physical models, using commercial LiDAR as the equipment is still a very challenging problem. However, the recent work of Zhu \etal~showed that by performing deep learning to complete feature extraction, LiDAR can be used to complete high-quality and robust 3D NLOS reconstruction\cite{zhuFastNonlineofsightImaging2021a}.

Aittala \etal~regarded the measurement data as the product of the hidden object and the light transport matrix\cite{aittalaComputationalMirrorsBlind2019} (as shown in Eq.~(\ref{eq2_1})). Exploiting deep neural models to generate matrices satisfied matrix decomposition completes passive NLOS imaging based on unsupervised learning. 
\cite{shenNonlineofsightImagingNeural2021} used unsupervised learning to complete active NLOS imaging. Specifically, \cite{shenNonlineofsightImagingNeural2021} exploited multilayer perceptron layers (MLP) to construct a neural transient field that maps measurement data into a hidden scene, and utilized a physical model to constrain the results, thereby obtaining superior reconstruction results.
Technically, such unsupervised learning methods~\cite{aittalaComputationalMirrorsBlind2019,shenNonlineofsightImagingNeural2021} are not data-driven methods, but the idea of using deep neural networks to simulate matrix factorization is instructive.

\bookmark[dest=\HyperLocalCurrentHref,level=2]{Challenges and Prospects}
\subsection{Challenges and Prospects}
NLOS imaging based on deep learning still faces many challenges, including lack of datasets, lockstep network structure, and limited generalization capability. Here, we discuss these three challenges and the corresponding prospects separately.

\bookmark[dest=\HyperLocalCurrentHref,level=3]{Dataset}
\subsubsection{Dataset}
Data-driven algorithms' performance largely depends on the dataset's quality (such as scale and reliability). However, for transient NLOS imaging, acquiring a sample of data requires a raster scan of the imaging area, which takes about 1 to 5 minutes~\cite{otooleConfocalNonlineofsightImaging2018,isogawaEfficientNonLineofSightImaging}. Therefore, constantly changing hidden objects and collecting tens of thousands of measurements to form a large-scale dataset is extremely difficult. For NLOS imaging scenes based on traditional cameras, although data collection is less time-consuming, it is still complicated to continuously change scenes and construct large-scale dataset because data augmentation operations such as crop and flip are not suitable for NLOS imaging.

Most existing works construct a rendering model to synthesize datasets for training, i.e., they first use public datasets in other fields as the hidden object (ground truth) and then synthesize measurement data through a rendering model. For transient NLOS imaging, the rendering model includes two parts: transient rendering and sensor model~\cite{chen_learned_2020,chopiteDeepNonLineofSightReconstruction2020}. The SPAD-based photon calculation model has been studied in detail, and they take into account factors such as noise, crosstalk and afterpulsing~\cite{hernandez_computational_2017}. For other types of NLOS imaging, Metzler \etal~used Berkeley Segmentation Dataset 500~\cite{martin_database_2001} as a hidden scene to calculate autocorrelation, which is then used to train the phase recovery network~\cite{metzlerDeepinverseCorrelographyRealtime2020}. For passive NLOS imaging using traditional cameras and ambient light, diffuse surface illumination models, such as the Phong Model~\cite{phongIlluminationComputerGenerated1975}, can be used for rendering to obtain synthesized passive NLOS data~\cite{zhouNonlineofsightImagingPhong2020}. 

Due to the sensitivity of deep learning algorithms to abnormal data, improving the accuracy of rendering models is extremely important to data-driven NLOS imaging methods. Another alternative is designing an automated/real-time data collection system to collect data in the experimental setup.

\begin{figure*}[!h]
    \centering
    \includegraphics[]{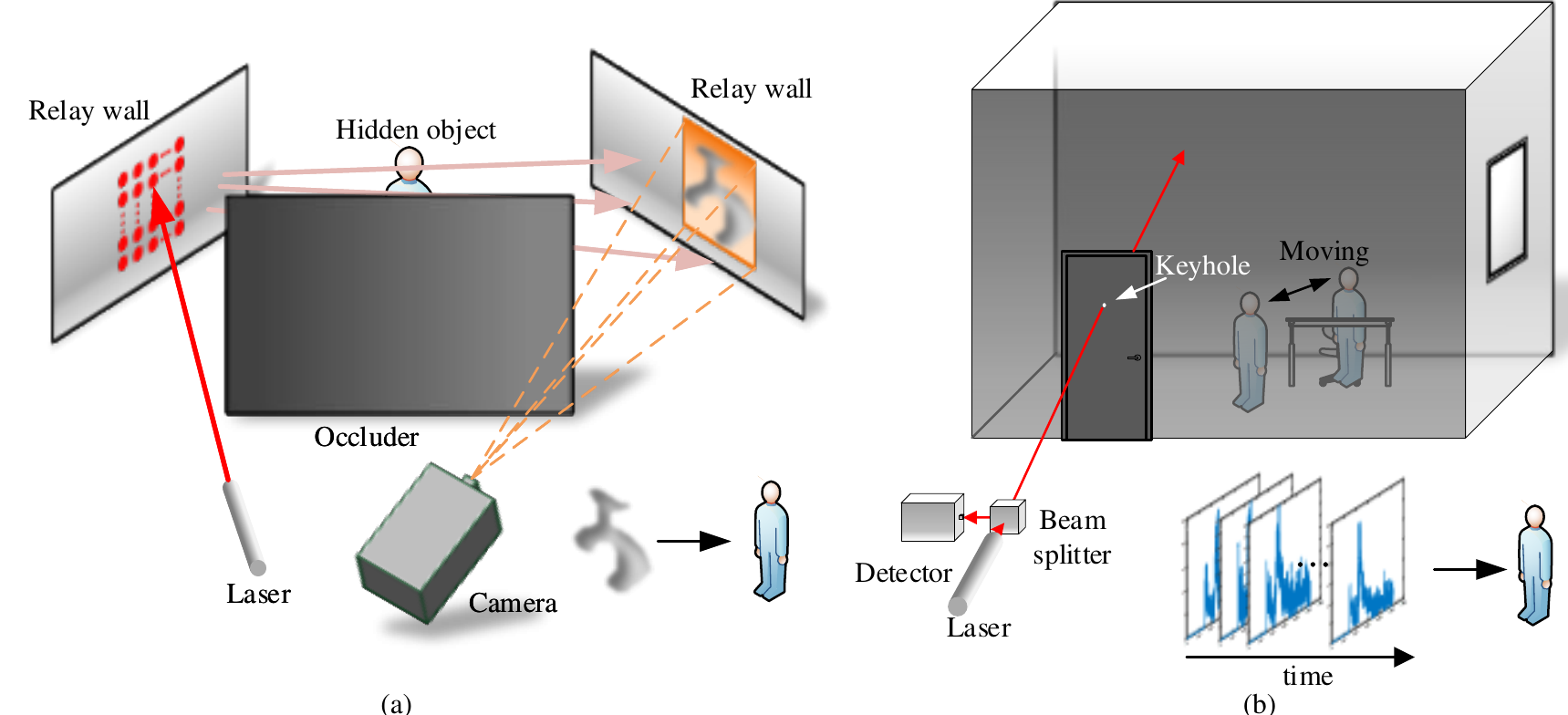}%
    \caption{New NLOS scenes. (a) Two bounce NLOS imaging. (b) Keyhole NLOS imaging. }
    \label{fig:newNLOS}
\end{figure*}

\bookmark[dest=\HyperLocalCurrentHref,level=3]{Network Architecture}
\subsubsection{Network Architecture}
According to Sec. IV~\ref{endToEnd} and \ref{combined}, most NLOS imaging networks currently use the classic network structure (or with tiny changes), such as U-Net~\cite{ronnebergerUnetConvolutionalNetworks2015a} and ResNet~\cite{heDeepResidualLearning2016a}. However, these well-known network structures are not necessarily suitable for NLOS imaging. Compared with recent works in other inverse problems, such as low-light image recovery~\cite{liu_pd-gan_2021} and image deblurring~\cite{rozumnyi_defmo_2020}, networks designed for specific tasks achieve better results. Therefore, it can be expected that in the future, more novel network architectures based on the characteristics of NLOS imaging(e.g., \cite{chen_learned_2020}) will be proposed to improve the performance of NLOS imaging.

\bookmark[dest=\HyperLocalCurrentHref,level=3]{Generalization}
\subsubsection{Generalization}
Limited by the dataset and network structure discussed above, current data-driven NLOS imaging algorithms have limited generalization capabilities. \cite{chopiteDeepNonLineofSightReconstruction2020} and \cite{chen_learned_2020} can get good test results on the public NLOS data~\cite{galindoDatasetBenchmarkingTimeresolved2019,otooleConfocalNonlineofsightImaging2018} through training on the synthetic training set, which means that they have a certain generalization ability. However, this does not mean that the generalization capabilities of current algorithms are sufficient. Faced with new parameters (such as BRDF, time jitter, aperture size) and new hidden scenes that do not exist in the training dataset, the trained network may not necessarily achieve acceptable results.

The most two direct ways to increase the generalization ability are improving the dataset's quality and the network structure, as discussed above. Another method is to use transfer learning so that a network trained on one kind of data can be quickly transferred to another kind of data. Besides, unsupervised learning is also a promising method. It does not rely on large amounts of data, but can use deep neural networks' powerful representation capabilities to train a model that meets the physical constraints (e.g., the optical transport matrix in~\cite{aittalaComputationalMirrorsBlind2019}), which has more substantial generalization capabilities.

\bookmark[dest=\HyperLocalCurrentHref,level=1]{New NLOS Scenes}
\section{New NLOS Scenes} \label{sec5}
In Sec.~\ref{sec2}-\ref{sec4}, most existing NLOS imaging studies have been introduced. These studies are accomplished in active and passive NLOS scenarios, as shown in Fig.~\ref{fig:introFig}. However, NLOS imaging is not limited to these scenes, and more new types of scenes are waiting to be developed. In this section, we take the recent two bounce NLOS imaging~\cite{vedaldi_imaging_2020} and keyhole NLOS imaging~\cite{metzler_keyhole_2021} as examples to introduce potential new NLOS scenes and their prospects.

\bookmark[dest=\HyperLocalCurrentHref,level=2]{Two-bounce NLOS imaging}
\subsection{Two bounce NLOS imaging}
Henley \etal~introduced the novel "two-bounce NLOS imaging", as shown in Fig.~\ref{fig:newNLOS}-(a)~\cite{vedaldi_imaging_2020}. In this new scene, the hidden object is blocked by an occluder, but there are two diffuse reflection surfaces $W_1$ and $W_2$ on both sides of the hidden object. The observer can reconstruct the hidden scene by scanning the laser on one of the diffuse reflection surfaces (e.g., $W_1$) and using a traditional camera to capture the two-bounce light on the other reflection surface (e.g., $W_2$). In other NLOS imaging, hidden objects play a role in reflecting light. However, in two bounce NLOS imaging, hidden objects play a role in blocking light. By capturing the projection of light on $W_2$ after being blocked by a hidden object, whether there is a hidden object point in each voxel can be judged.

As discussed in~\cite{vedaldi_imaging_2020}, two bounce NLOS imaging can be applied to seeing behind trucks for autonomous vehicles and imaging between windows for search and rescue. These applications are extensions to traditional NLOS imaging (i.e., three-bounce NLOS imaging or passive NLOS imaging) applications and illustrate the broad application prospects of ``turning walls into mirrors''.

\bookmark[dest=\HyperLocalCurrentHref,level=2]{Keyhole Imaging}
\subsection{Keyhole Imaging}
For all the NLOS imaging systems introduced above, there is a limitation that a large illumination/imaging area is required.
However, in some NLOS scenes (e.g., hidden objects are blocked in a room with curtains), large imaging areas are not allowed, and only a few small holes can be used. 
Metzler~\etal~used a beam splitter to propose an imaging system similar to confocal NLOS but does not require scanning, called keyhole imaging, as shown in Fig.~\ref{fig:newNLOS}-(b)~\cite{metzler_keyhole_2021}. In this system, the pulsed laser and the time-resolved detector only use a small hole to illuminate and detect the hidden space. When the object in the hidden space moves with time, expectation-maximization can be used to complete the task of NLOS shape reconstruction and localization.

\bookmark[dest=\HyperLocalCurrentHref, level=1]{Conclusion}
\section{Conclusion} \label{sec6}
In this paper, we reviewed the existing NLOS imaging techniques with different illumination conditions. Besides, we also discussed the recent data-driven reconstruction algorithm and introduced some new types of NLOS scenes. All these works demonstrated that NLOS imaging can significantly improve imaging equipment's view and capabilities, allowing hidden objects to be seen. There are still several challenges that need to be addressed in the future work, including how to improve the SNR and reduce the ill-posedness of the problem, how to use more effective prior information, and how to reduce scanning time, making NLOS imaging technology more practical.


%




\ifCLASSOPTIONcaptionsoff
  \newpage
\fi



%


\bibliographystyle{IEEEtran}
\bibliography{egbib}

%








\end{document}